\documentclass[journal]{IEEEtran}

\usepackage{xcolor,soul,framed,caption} 

\colorlet{shadecolor}{yellow}
\usepackage[pdftex]{graphicx}
\graphicspath{{../pdf/}{../jpeg/}}
\DeclareGraphicsExtensions{.pdf,.jpeg,.png}

\usepackage[cmex10]{amsmath}
\usepackage{array}
\usepackage{mdwmath}
\usepackage{mdwtab}
\usepackage{eqparbox}
\usepackage{url}
\usepackage{supertabular}                                             
\usepackage{amsmath,amssymb}
\usepackage{caption}
\usepackage{subcaption}
\usepackage{float}
\usepackage{booktabs}
\usepackage{tabularray}
\usepackage{multirow}
\usepackage{graphicx}
\usepackage[ruled, lined, longend, linesnumbered]{algorithm2e}
\usepackage[square, numbers, comma, sort&compress]{natbib}  
\usepackage{verbatim}  
\usepackage{vector}  

\usepackage{listings}
\usepackage{xcolor}

\definecolor{codegreen}{rgb}{0,0.6,0}
\definecolor{codegray}{rgb}{0.5,0.5,0.5}
\definecolor{codepurple}{rgb}{0.58,0,0.82}
\definecolor{backcolour}{rgb}{0.95,0.95,0.92}

\lstdefinestyle{mystyle}{
    backgroundcolor=\color{backcolour},   
    commentstyle=\color{codegreen},
    keywordstyle=\color{magenta},
    numberstyle=\tiny\color{codegray},
    stringstyle=\color{codepurple},
    basicstyle=\ttfamily\footnotesize,
    breakatwhitespace=false,         
    breaklines=true,                 
    captionpos=b,                    
    keepspaces=true,                 
    numbers=left,                    
    numbersep=5pt,                  
    showspaces=false,                
    showstringspaces=false,
    showtabs=false,                  
    tabsize=2
}

\usepackage{multirow}
\usepackage{multicol}
\usepackage{graphicx}
\usepackage{svg}
\usepackage{hyperref}

\usepackage{amssymb}
\usepackage{pifont}
\newcommand{\cmark}{\ding{51}}%
\newcommand{\xmark}{\ding{55}}%

\begin{document}
\bstctlcite{IEEEexample:BSTcontrol}
    \title{FLight: A Lightweight Federated Learning Framework in Edge and Fog Computing}
  \author{Wuji Zhu, Mohammad Goudarzi, and Rajkumar Buyya

  \thanks{Wuji Zhu and Rajkumar Buyya are with the Cloud Computing and Distributed Systems (CLOUDS) Laboratory, School of Computing and Information Systems, The University of Melbourne, Australia (e-mail: wujiz1@student.unimelb.edu.au, rbuyya@unimelb.edu.au).}
   \thanks{Mohammad Goudarzi is with the School of Computer Science and Engineering, The University of New South Wales (UNSW), Australia (email: m.goudarzi@unsw.edu.au)}
}


\maketitle

\begin{abstract}
The number of Internet of Things (IoT) applications, especially latency-sensitive ones, have been significantly increased. So, Cloud computing, as one of the main enablers of the IoT that offers centralized services, cannot solely satisfy the requirements of IoT applications. Edge/Fog computing, as a distributed computing paradigm, processes, and stores IoT data at the edge of the network, offering low latency, reduced network traffic, and higher bandwidth. The Edge/Fog resources are often less powerful compared to Cloud, and IoT data is dispersed among many geo-distributed servers. Hence, Federated Learning (FL), which is a machine learning approach that enables multiple distributed servers to collaborate on building models without exchanging the raw data, is well-suited to Edge/Fog computing environments, where data privacy is of paramount importance. Besides, to manage different FL tasks on Edge/Fog computing environments, a lightweight resource management framework is required to manage different incoming FL tasks while does not incur significant overhead on the system. Accordingly, in this paper, we propose a lightweight FL framework, called FLight, to be deployed on a diverse range of devices, ranging from resource-limited Edge/Fog devices to powerful Cloud servers. FLight is implemented based on the FogBus2 framework, which is a containerized distributed resource management framework. Moreover, FLight integrates both synchronous and asynchronous models of FL. Besides, we propose a lightweight heuristic-based worker selection algorithm to select a suitable set of available workers to participate in the training step to obtain higher training time efficiency. The obtained results demonstrate the efficiency of the FLight. The worker selection technique reduces the training time of reaching 80\% accuracy by 34\% compared to sequential training, while asynchronous one helps to improve synchronous FL training time by 64\%.  
\end{abstract}


%
\IEEEpeerreviewmaketitle


\section{Introduction}
\IEEEPARstart{R}{ecently}, Machine Learning (ML) applications such as speech recognition, natural language processing, computer vision, decision-making, and recommendation systems have gained significant popularity. An essential factor for the successful development and deployment of ML applications and models is accessing a large amount of data \cite{ORI_4,goudarzi2022cloud}. Due to the ever-increasing number of Internet of Things (IoT) applications, exploding amounts of data have been continuously being generated from different sources. However, the increasing awareness of data protection and privacy concerns prevent unconstrained data collection and access. General Data Protection Regulation (GDPR) \cite{ORI_7} was introduced in 2018, which regulates and formalizes how organizations should utilize collected data. Similarly, the United States has the California Consumer Privacy Act (CCPA) \cite{ORI_8}, which also legislates practices of using customer data. Also, even in scenarios where data collection is legally permitted, data silo issues still may happen, referring to the cases in which a data repository controlled by a department (e.g., hospitals or organizations) is isolated and incapable of reciprocal operation. Thus, collecting huge amount of raw data from different sources for ML applications and models face an important barrier. Accordingly, a method to access a rich amount of available data in distributed sites while satisfying data access, protection, and privacy constraints is required. Federated Learning (FL) arises as a promising method to train models based on data from different sites while satisfying data protection rules \cite{bonawitz2019towards}.
\par
The main idea of FL is to communicate the model weights or parameters rather than the raw data for training models. In FL, models are first trained locally over multiple devices on different sites. Because the data are trained locally, data privacy is protected, and there is no violation of the GDPR or organizations' rules. After a few epochs of local training performed at different sites, model weights are extracted and transmitted to an aggregation server. Afterward, the aggregation server merges the received weights from different sites using different algorithms to build a more robust model. So, FL can solve the data silo issues by providing access to more data for ML applications and models. FL features not only leads to more robust models but also promise better data security, privacy, and also more efficient data transmission \cite{ORI_11,ORI_12}.
\par
The rapid increase in the number of IoT applications and their heterogeneous requirements (such as low latency, high privacy, and geo-distributed coverage) has caused centralized computing solutions, such as Cloud Computing, to fall short in solely providing efficient services for a wide variety of IoT applications \cite{goudarzi2020application,xu2019computation}. Consequently, Edge/Fog computing, as a distributed computing paradigm, has emerged, in which heterogeneous distributed servers in the proximity of end-users process and store the data \cite{goudarzi2022scheduling}. Since the resources in Edge/Fog computing are often limited compared to the Cloud resources, Edge/Fog servers may require collaboration with other Edge/Fog servers or Cloud resources for the proper execution of diverse IoT applications. In such a distributed computing environment, FL can be appropriately fit due to several reasons. First, Edge/Fog devices have access to the local data of many IoT applications, which can be used to train more powerful models. Second, privacy and data protection can be satisfied using the localized data in the Edge/fog resources. Third, each Edge/Fog resource often does not have sufficient resources and data for the training of powerful models, while using FL, they can share their data and resources to more efficiently train better models. Fourth, the heterogeneous nature of data in different Edge/Fog servers offers more valuable data for training better models. Fifth, FL helps resource-limited Edge/Fog servers, which have higher limitations in terms of resources and/or data, to more frequently update their models based on the model shared by the model aggregation servers. However, the heterogeneity in Edge/Fog computing environments requires further considerations for the successful deployment of FL.    
\par
Heterogeneity among servers participating in FL includes systems' resources, networking characteristics, available data size, and availability \cite{ORI_16}. Different Systems' resources comprise different CPU frequencies, CPU utilization rates, RAM, and GPU, just to mention a few. These diverse resources result in different required training times for each server to train a fixed amount of epochs with a predefined amount of data. Considering different computing times among various servers participating in FL, fast servers often need to wait for slower servers, which constrains the efficiency of the training model in a distributed manner \cite{ORI_16}. Different Network characteristics can also influence the deployment of FL \cite{ORI_17}. Because different servers have various networking capabilities (e.g., download and upload speeds, dropout rates), the time required to transmit the common model structure used by FL from an aggregation server to other servers will vary significantly \cite{ORI_17}. In extreme cases, other servers can finish multiple training rounds when a model is transmitted to servers with minimal network capacity. Such heterogeneous characteristics incur slower sites to wait for a long period, which is a waste of resources. Different data sizes for training a model on each server also challenge the FL. A large training data model can produce a much more robust and accurate ML model compared to servers with limited data. Thus, combining models of sites with various amounts of data will underperform compared to only combining models from sites with large amounts of data. However, simply dropping results from sites with limited data failed to explore the potential of all data available. Hence, to successfully conduct FL, it is essential to address the heterogeneity among different participants of the training process. To address these challenges, intelligent decisions about participant selection for each round of training, network allocation for each participant, and merging algorithms should be carefully investigated \cite{ORI_16, ORI_17}. 
\par
Although some works attempt to improve time efficiency or computational energy efficiency based on system statistics, none of them have attempted to integrate asynchronous FL. Furthermore, to the best of our knowledge, only a few works that optimized the deployment of FL mechanisms provide the source code of their research. Besides, among the studies that offer their source code, it is very difficult to implement new worker selection policies or new models.
To address these challenges and limitations, we propose the containerized FLight framework based on the FogBus2 resource management framework. The main contributions of this paper are:   
\begin{itemize}
\item Designing and implementing a containerized FL system by extending the FogBus2 resource management framework.

\item  Implementing a lightweight mechanism so that participants' selection policies for FL and employed algorithms for merging models from different participants can be efficiently integrated.

\item Implementation of an asynchronous participants selection policy based on real data obtained from FogBus2 and underlying computational resources. 

\item The practical implementation and analysis demonstrating Flight frameworks and their integrated policies outperform synchronous and other baseline techniques in terms of time efficiency in reaching the desirable accuracy.
\end{itemize}

The rest of the paper is organized as follows. Section \ref{related_work} discusses the relevant literature. Section \ref{FL_model_design} presents the design of the Flight framework, worker selection, and respective problem formulation. Section \ref{performance_evaluation} describes the experimental setup and evaluates the performance of the Flight framework. Finally, Section \ref{conclusion} concludes the paper and draws future work.
\section{Background and Related Work}
\label{related_work}
In this section, we discuss the required FL concepts, review techniques targeting to optimize FL based on system parameters, and describe FogBus2's main components. The terminologies used for FL are summarized in table \ref{table:abbreviation}.
\begin{table*}[]
\caption{Abbreviations}
\label{table:abbreviation}
\resizebox{\textwidth}{!}{%
\begin{tabular}{|l|l|l|}
\hline
Terminology & Explanation & Abbreviation \\ \hline
Aggregation Server & Server used to aggregate models from different parties & $AS$ \\ \hline
Worker & Server contributing to FL via pushing the local model to aggregation servers & $w$ \\ \hline
Aggregation algorithms & Algorithms used to combine model results from different workers to a single model & $f_{aggr}$ \\ \hline
Worker selection algorithm & Algorithms used by aggregation server to select workers for participating in FL. & $f_{sel}$ \\ \hline
Aggregation server model weights & The model weights of the model held by aggregation server (aggregate for $i$ times) & $M_{asi}$ \\ \hline
Worker model weights & The weights of the model locally held by worker $x$ (based on version $i$ of the aggregation server and trained for j times) & $Mw_{x, i, j}$ \\ \hline
Worker averaging weights & Weights allocated for weighted federated averaging for worker $x$ & $WEI_x$ \\ \hline
\end{tabular}%
}
\end{table*}
\subsection{FL Concepts}
FL is an approach to train a shared ML model based on data from multiple parties \cite{ORI_9}.  A server can take different roles \cite{ORI_23, ORI_24}. Firstly, the worker, which is a server that contributes to the FL by training a commonly agreed model by their local data. Secondly, the aggregation server, which is a server collecting model weights from workers and aggregating them via different aggregation algorithms. Lastly, peers refer to servers communicating with each other, while none of them are central aggregation servers.
\par
FL is featured by pushing model parameters periodically between an aggregation server for aggregating model weights and workers. A complete FL process is partitioned into three stages: 1) Connection Establishment, 2) Model Training, and 3) Training Evaluation \cite{ORI_9,ORI_23}. As FL is a collaborative process, the connection needs to be established before the start of FL training. In the connection establishment stage, different servers identify their roles. Also, the common ML model that each server uses for FL is set. The aggregation server sends the description of the ML model to each worker. In the model training stage, the aggregation server selects workers to participate in FL training. This process is referred to as worker selection. Worker selection is a trade-off between efficiency and accuracy. Intuitively, selecting more workers will allow the FL to access more data, leading to better model performance. However, this means faster workers must wait for slower workers, requiring more training time to reach the desired accuracy \cite{ORI_25}. After worker selection, the aggregation server informs workers to start the FL training. Then, workers download the version of the model that the aggregation server provides. There are two methods to achieve this. One is an aggregation server sending information about the model directly to workers. This method is easy to implement and straightforward. However, this will cause network congestion at the aggregation server point \cite{ORI_26}. At the same time, the worker may not be available to receive the model due to network availability at the time the aggregation server tries to send the model. An alternative method is the aggregation server to push the ML model to a database and provides download credentials to all selected workers \cite{ORI_9}, \cite{ORI_26}. Although it is harder to implement, this method is more friendly to workers with various network conditions. This is because workers can download the model at their desired time. At the same time, the aggregation server only needs to communicate with the database instead of multiple workers, which relieves network pressure. After workers finish training for a few epochs, they send the model weights back to the aggregation server. Next, the aggregation server merges received model weights from workers. After the aggregation server finishes averaging, one epoch of FL training is finished. In the training evaluation stage, the aggregation server evaluates the performance of the averaged model to decide whether to repeat the FL training step or not. The evaluation process is usually performed via two methods. The first method is the aggregation server uses locally available data to test the performance of the aggregated model. Alternatively, the aggregation server asks workers to download the latest model, evaluate the model performance locally, and then provides an average accuracy \cite{ORI_9}. If accuracy does not meet the expectation, stages two and three are executed repeatedly.
\par
The aggregation server aggregates model weights once a sufficient number of workers finish transmitting their models to its cache. However, this step does not require the aggregation server to wait until all selected workers respond, resulting in the asynchronous situation \cite{ORI_9, ORI_31, ORI_32}. There are three possible cases regarding worker $W$ sending the model weights trained by local data and the aggregation server $AS$ starting merging model weights from workers: 1) Model of $W$ arrives before $AS$ starts to aggregate model weights from workers, and when $AS$ starts to aggregate model weights from workers. It includes model weights $W$ in the aggregation process. 2) Model of $W$ arrives after $AS$ starts to aggregate model weights from workers, and $AS$ refuses to receive model weights of $W$ even when $W$ responds to $AS$. Model weights of $W$ will not be included in the aggregation process. 3) The $W$ model arrives after $AS$ starts to aggregate model weights from workers. Although $AS$ does not include model weights of $W$ for the current round of aggregation, $AS$ receives the model and uses it for the next round of aggregation. The combination of points one and two is called synchronous FL and otherwise asynchronous FL. The advantage of synchronous mode is the simplicity of merging algorithms since all worker model weights are based on the same version of aggregation server model weights. In contrast, model weights may be based on different versions of aggregation server weights for asynchronous FL. Thus extra consideration is necessary for the merging algorithms. The advantage of asynchronous FL is time efficiency compared to the synchronous mode since the aggregation server does not have to wait for slower workers before aggregation and to proceed to the next epoch.
\par
Aggregation methods are essential for the successful update of the aggregation server model, affecting the training time required to reach a desirable accuracy that depicts the importance of understanding aggregation algorithms (e.g., federated averaging, linear weighted averaging, polynomial weighted averaging, exponential weighted averaging). Aggregation algorithms that are biased to the response from workers with more training data or newer versions of the aggregation server model diminish the negative effects of less active workers. In contrast, algorithms with little or no bias risk being negatively influenced in terms of aggregated model performance by low-performing workers. Thus, a trade-off between these two kinds of methods is required, and difference aggregation algorithms need to be examined.
\subsection{Related Work}
In this section, recent research to improve the efficiency of FL based on system parameters with a focus on tuning aggregation frequency, worker selection algorithms, and training epoch tuning are studied.
\par
Tuning aggregation frequency is a method to adjust the epoch number that workers require to train locally before responding to the aggregation server. First attempts to optimize FL based on system parameters targeted to quantify several system resources \cite{ORI_42}. The main optimization goal of this work is to achieve maximum accuracy under a fixed resource budget. This is achieved by tuning the frequency of global aggregation. Since the resource budget is fixed, the total training epochs a worker can perform are fixed. Thus, tuning the frequency of global aggregation can also determine the number of aggregations performed. While the evaluation results depict better accuracy by a few percent, there are some drawbacks to such methods. Firstly, the technique failed to consider the asynchronous case, and the budget is calculated as the minimum budget among all workers, which ignores exploring all resources from other workers with more substantial computation power. Secondly, it intends to exhaust all available resources to achieve the best accuracy, resulting in workers with rich resources to calculate for a large number of epochs before responding to the aggregation server, leading to training time inefficiency. Lastly, the technique failed to adaptively adjust based on workers' performance.
\par
Several research studies in the literature aim at optimizing the worker selection policy. These works intend to improve the energy efficiency or training time of the FL process by selecting appropriate workers to balance resource efficiency and model accuracy. The authors of \cite{ORI_43} proposed a heuristic-based worker selection policy. This work converts all quantified workers' resources to the time required to transmit and train the model. It assumes all workers can only communicate the model sequentially with the aggregation server and set up the total time allowed between each round of aggregation as a variable. Next, the worker selection policy continuously adds a worker to the worker set until the required time to finish training and transmit the model for all workers exceeds the total time allowed for the current round of aggregation. If the model accuracy increases slowly, then the total time allowed for the current epoch increases to allow more workers to participate in FL training. The technique does not support asynchronous mode. Thus, the issue of fast computing workers waiting for slow workers still exists. Also, although the technique assumes all workers communicate the model weights with the aggregation server sequentially, which simplifies the optimization model, it ignores the potential of saving time by transmitting the model in parallel. While \cite{ORI_42, ORI_43} tried to select as many workers as possible within a given resource budget, the \cite{ORI_45} used a Reinforcement Learning (RL) for the worker selection process \cite{ORI_45}. This work considers the energy cost, which is a trade-off between the maximum number of selected workers against the energy consumption \cite{ORI_45}. The RL method considers the cost of each step as the total energy consumed by a set of workers to perform the training based on CPU power and the required time for processing. At the same time, the reward is calculated as the accuracy improvement of the aggregated model. The employed RL technique offers an adaptability feature for this work. However, this research only focuses on the trade-off between energy consumption and model accuracy, while time efficiency is not considered. Another similar research considered energy consumption while targeting the trade-off between accuracy and energy consumption \cite{ORI_46}. The above-mentioned works \cite{ORI_43, ORI_45, ORI_46} primarily depend on the resources participating in the FL process by each worker. They generally assume that more workers will allow the model to be exposed to more data; hence, it improves the accuracy of the shared model. In this way, worker selection policies are more biased towards faster workers and assume more epochs of training conducted by faster workers can result in better accuracy compared to fewer epochs of training that slower workers can do. However, another research suggests that the direction of model update gradients indicates the effectiveness and contribution of different workers \cite{ORI_49}. This technique requests all workers to train for a few epochs and fetch their response rate. Next, the algorithm computes the average model weights in the first round of aggregation. Then, the algorithm compares all model response rates with the averaged model by calculating the norm difference \cite{ORI_49}. A more significant norm difference indicates that the model update direction is different from the majority's trend, suggesting the corresponding worker has little or no contribution to the overall accuracy. On the other hand, workers with minimum norm difference with the averaged model indicate the model's effectiveness in improving the accuracy of the FL process \cite{ORI_49}. This technique conducts a worker selection process more biased towards workers with little norm compared to the averaged model. While this technique plan to save time for FL training by reducing training epochs, more training time is required since slower workers will be selected over faster ones if they have less model norm difference. Previous worker selection strategies have the idea of allowing fast computing workers to participate in FL training and then adding slow workers progressively in later rounds. However, this approach has the drawback that slow workers cannot communicate with the aggregation server for a long time and can only participate in the FL in the last few rounds of training. Accordingly, the authors in \cite{ORI_51} considered the time since the last time a worker fetched model weights from the aggregation server and contributed by responding to the updated model. The integrated ranking formula for worker selection considers such time until the last contribution, together with other factors like computation time and energy consumption. Then, the top workers will be selected, leading to the selection of slower workers before faster workers if they have not communicated with the aggregation server for a long time. While the overall time increases in this technique, the convergence rate is faster. However, because selecting too many slow workers can negatively affect the training time, \cite{ORI_52} proposed a worker selection technique through clustering to control the number of slow workers in the FL training.
\par
Most recent research attempts to accelerate the FL process by allocating tasks with different workloads to workers with different capacities to allow heterogeneous workers to finish tasks simultaneously \cite{ORI_44, ORI_47, ORI_48, ORI_49, ORI_50, ORI_51, ORI_52}. A joint optimization algorithm is proposed by \cite{ORI_44}, in which the transmission power of workers and aggregation servers are adjusted to allow transmission efficiency. Instead of tuning the transmission rate to achieve the balance of worker time consumption through tuning the bandwidth, \cite{ORI_47} intends to tune the amount of data utilized by different workers so that slow and fast workers can finish the training simultaneously. The \cite{ORI_48} proposed to use the dropout method to achieve similar computation time among workers rather than tuning the amount of data. The authors of \cite{ORI_50} proposed the concept of offloading part of computation tasks to the aggregation server or nearby servers to reduce the burden of slower workers. While this method is more efficient in terms of training time and resource usage, it requires special considerations as the data should be forwarded from the workers to other servers.
\subsubsection{Qualitative Comparison}
\begin{table*}
\centering
\caption{Recent research on system parameter-based federated learning optimisation}
\label{table:big}
\resizebox{1.05\textwidth}{!}{%
\begin{tblr}{
  row{even} = {c},
  row{3} = {c},
  row{5} = {c},
  row{7} = {c},
  row{9} = {c},
  row{11} = {c},
  row{13} = {c},
  cell{1}{1} = {r=2}{c},
  cell{1}{2} = {r=2}{c},
  cell{1}{3} = {r=2}{c},
  cell{1}{4} = {r=2}{c},
  cell{1}{5} = {r=2}{c},
  cell{1}{6} = {r=2}{c},
  cell{1}{7} = {r=2}{c},
  cell{1}{8} = {c=7}{c},
  cell{1}{15} = {c=5}{c},
  cell{1}{20} = {c=6}{c},
  vline{1-7} = {1}{},
  vline{8-15} = {1}{},
  vline{15-20} = {1}{},
  vline{20-26} = {1}{},
  vline{1-8} = {2}{},
  vline{8-26} = {2}{},
  vline{1-26} = {3-14}{},
  hline{1,3-15} = {1-25}{},
  hline{2} = {8-25}{},
}
Work   & Implementation & {Updt\\Freq } & {Perf\\Check } & {Time\\Const } & Asyn & WSP & System Paramters &     &     &     &     &     &     & Tuned FL Parameters &     &     &    &     & Derived Parameters &     &     &     &     &     &  \\
       &                &               &                &                &      &     & F                & P   & B   & G   & D   & WDS & WL  & T                   & t   & MDR & OR & TDS & TC                 & TU  & TW  & EC  & EU  & L   &  \\
\cite{ORI_43}     & Analytical     & Epoch         & \xmark             & \cmark            & \xmark   & \cmark & \cmark              & \xmark  & \cmark & \xmark  & \xmark  & \cmark & \xmark  & \xmark                  & \xmark  & \xmark  & \xmark & \xmark  & \cmark                & \cmark & \cmark & \xmark  & \xmark  & \xmark  &  \\
\cite{ORI_46}     & Analytical     & Epoch         & \xmark             & \cmark            & \xmark   & \cmark & \cmark              & \cmark & \cmark & \cmark & \xmark  & \cmark & \cmark & \xmark                  & \xmark  & \xmark  & \xmark & \xmark  & \cmark                & \cmark & \cmark & \cmark & \cmark & \cmark &  \\
\cite{ORI_52}     & Analytical     & Epoch         & \xmark             & \cmark            & \xmark   & \cmark & \cmark              & \cmark & \cmark & \cmark & \cmark & \cmark & \cmark & \xmark                  & \xmark  & \xmark  & \xmark & \xmark  & \cmark                & \cmark & \xmark  & \xmark  & \xmark  & \xmark  &  \\
\cite{ORI_44}     & Simulation     & Once          & \xmark             & \cmark            & \xmark   & \cmark & \cmark              & \cmark & \cmark & \cmark & \cmark & \xmark  & \cmark & \xmark                  & \xmark  & \xmark  & \xmark & \xmark  & \xmark                 & \cmark & \xmark  & \cmark & \cmark & \cmark &  \\
\cite{ORI_45}     & Simulation     & Epoch         & \cmark            & \xmark             & \xmark   & \cmark & \cmark              & \xmark  & \cmark & \xmark  & \cmark & \cmark & \xmark  & \xmark                  & \xmark  & \xmark  & \xmark & \xmark  & \cmark                & \cmark & \cmark & \xmark  & \xmark  & \xmark  &  \\
\cite{ORI_48}     & Simulation     & Epoch         & \xmark             & \xmark             & \xmark   & \xmark  & \cmark              & \cmark & \cmark & \xmark  & \xmark  & \cmark & \xmark  & \xmark                  & \xmark  & \cmark & \xmark & \xmark  & \cmark                & \cmark & \xmark  & \xmark  & \xmark  & \xmark  &  \\
\cite{ORI_49}     & Simulation     & Epoch         & \xmark             & \cmark            & \xmark   & \cmark & \cmark              & \cmark & \cmark & \cmark & \cmark & \xmark  & \cmark & \xmark                  & \xmark  & \xmark  & \xmark & \xmark  & \cmark                & \cmark & \cmark & \cmark & \cmark & \xmark  &  \\
\cite{ORI_42}     & Prototype      & Epoch         & \cmark            & \cmark            & \xmark   & \xmark  & \cmark              & \cmark & \cmark & \xmark  & \xmark  & \xmark  & \cmark & \cmark                 & \cmark & \xmark  & \xmark & \xmark  & \xmark                 & \xmark  & \xmark  & \xmark  & \xmark  & \xmark  &  \\
\cite{ORI_47}     & Prototype      & Epoch         & \xmark             & \cmark            & \xmark   & \xmark  & \cmark              & \xmark  & \cmark & \xmark  & \xmark  & \cmark & \xmark  & \cmark                 & \cmark & \xmark  & \xmark & \cmark & \cmark                & \cmark & \cmark & \cmark & \xmark  & \xmark  &  \\
\cite{ORI_50}     & Prototype      & Once          & \xmark             & \cmark            & \xmark   & \xmark  & \cmark              & \xmark  & \xmark  & \xmark  & \xmark  & \xmark  & \cmark & \xmark                  & \xmark  & \xmark  & \xmark & \xmark  & \cmark                & \cmark & \xmark  & \xmark  & \xmark  & \xmark  &  \\
\cite{ORI_51}     & Prototype      & Epoch         & \cmark            & \xmark             & \xmark   & \cmark & \cmark              & \xmark  & \cmark & \xmark  & \xmark  & \cmark & \xmark  & \xmark                  & \cmark & \xmark  & \xmark & \xmark  & \xmark                 & \xmark  & \xmark  & \xmark  & \xmark  & \cmark &  \\
FLight & Prototype      & Epoch         & \cmark            & \cmark            & \cmark  & \cmark & \cmark              & \cmark & \cmark & \xmark  & \xmark  & \cmark & \cmark & \cmark                 & \cmark & \xmark  & \xmark & \cmark & \cmark                & \cmark & \cmark & \xmark  & \xmark  & \cmark &  
\end{tblr}
}
\end{table*}

This section identifies important parameters in FL and summarizes the properties of the current literature considering the identified parameters. Table \ref{table:big} depicts an overview of the current FL studies and their properties. In what follows, the identified parameters and their respective symbols in the table are explained.
\begin{itemize}
\item Implementation level: A prototype system refers to the technique that is deployed on different servers, and the results are practically verified. Analytical refers to only mathematically modeling the FL training process, such that the training time and accuracy of the underlying model are not tested on any data but tested by mathematical models. Simulation mimics the network delay and computation difference on a single server.
\item Update Frequency (Updt Freq): It refers to the frequency of updates (i.e., optimization) in each technique. While each epoch means the optimization strategy will be updated each time the aggregation server finishes aggregating model weights, once means the optimization policy is only calculated once and will keep steady throughout the FL training process.
\item Performance Check (Perf Check): It describes whether performance is analyzed throughout the FL training.
\item Time Constraint: It shows whether training time is considered a constraint throughout the training process.
\item Asynchronous (Asyn): It identifies if the technique can run asynchronously.
\item Worker Selection Policy (WSP): It identifies if the technique offers worker selection policy.
\item System parameters: It is the raw statistics extracted from a system used to formulate FL optimization algorithms. These statistics include: 1) F: processor frequency, 2) P: transmission power, 3) B: bandwidth, 4) G: channel gain, 5) D: aggregation server and worker distance, 6) WDS: worker data size, 7) WL: Workload required to compute one epoch of training from the worker side.
\item Tuned FL Parameters: 1) T: Total epoch number trained throughout the FL, 2) t: Number of epochs trained on workers between two aggregation processes from the aggregation server, 3) MDR: model dropout ratio, 4) TDS: training data size, 5) OR: offloading ratio.
\item Derived Parameter: 1) TC: the time required for workers to communicate model weights, 2) TU: the time required for training as well as loading the model, 3) TW: the time required to wait until an aggregation server or a worker is available, 4) EC: energy required for local training, 5) EU: energy required to upload the model, 6) L: training loss.
\end{itemize}
\par
Considering the current literature, each work depends on different system parameters and derives different intermediate results for optimization purposes. However, the current literature always compares their performance with no optimization or random worker selection policy \cite{ORI_42,ORI_43,ORI_44,ORI_45,ORI_46,ORI_47,ORI_48,ORI_49,ORI_50,ORI_51,ORI_52}, which partially proves the proposed algorithms' necessity. However, it is hard to compare these works against each other because reproducing the experimental results is cumbersome due to the difference between their system setup and programming style. Consequently, implementing an FL framework that allows different techniques to be integrated allows a fair comparison for studying the effectiveness of different techniques. Besides, none of the works has studied the asynchronous FL to utilize the full computational capacity of workers. While unselected workers have to wait until the next round of federated worker selection to contribute to the FL training, asynchronous FL permits all workers to participate in FL. This is because asynchronous FL is not concerned about slow workers to keep fast workers waiting. Instead, asynchronous FL can start aggregation when fast workers finish responding with their trained model weights. Also, when slow workers finish training, the aggregation server can merge the results from slow workers. Moreover, the worker selection policy is required to select the most suitable workers for the purpose of FL training.
\par
Accordingly, to address these challenges,  we design a lightweight FL framework called FLight by extending the FogBus2 resource management framework. Flight provides a mechanism to integrate different worker selection policies and also proposes a lightweight policy. Finally, FLight offers asynchronous FL to further improve the current literature.
\subsection{FogBus2 Framework}
To implement the FL framework based on system parameters, the above-mentioned system parameters should be provided for the FL on demand. Moreover, the FL framework requires access and manage distributed servers in the environment, which are often highly heterogeneous in terms of hardware, operating systems, and software. Besides, the connection establishment stage of the FL requires various interaction models and online information about the whole system. To satisfy these requirements, the FogBus2 framework \cite{deng2021fogbus2} is chosen as the underlying resource management framework.
\par
FogBus2 is a new Python-based framework consisting of five lightweight and containerized modules (called Master, Actor, User, Task Executor, and RemoteLogger) that support centralized, distributed, and hierarchical deployments. To suit different resource management requirements, it offers several mechanisms and associated policies, including registration, profiling, scheduling, scalability, dynamic resource discovery, IoT application integration, database integration, etc.  
\par
FogBus2 provides systems parameters required for the FL, such as CPU frequency, RAM, resource utilization, and networking characteristics, just to mention a few, using a profiling module integrated into its main components. It supports on-demand and periodical profiling, which is helpful for asynchronous FL that frequently merges model weights from workers. After each merging, new worker selection can be conducted, resulting in a more frequent worker selection process. Frequent updates about system parameters among workers allow asynchronous FL to update worker selection accordingly. A more dynamic and adaptive worker selection policy results in more time-efficient FL. This suggests that frequent system parameter update is beneficial for more efficient FL practice. Also,  systems parameters in the FogBus2 framework are accessible centrally and also in distributed databases, facilitating the process of obtaining system parameters.
\par
As FogBus2 is dockerized \cite{ORI_19}, it helps the FL framework to be easily deployed on machines with heterogeneous environments and various operating systems. FL intends to train ML models, requiring support from different ML libraries such as PyTorch. While manually installing ML dependencies on different operating systems require various configuration, containerized FogBus2 framework facilitates this process by adding required dependencies to the configuration file.
\par
Finally, FogBus2 provides a resource discovery sub-module that facilitates the connection establishment process of FL. The connection establishment process of FL occurs when different participants connect with each other and know their role in the FL framework. This requires network configuration to allow different servers to communicate with each other while message handlers on each server need to forward the message to the FL training agent. Moreover, it requires the participant who starts the FL process to be aware of potentially available computing resources within the network, so it can send initiative commands and invite them to participate in FL. Moreover, it has integrated task allocation policies, allowing all available sites with sufficient computing resources to conduct FL training, which is convenient for testing and verifying FL implementation.
\section{Flight: A Lightweight FL Framework}
\label{FL_model_design}
In this section, we describe the FLight framework and how it extends the FogBus2 framework.
\subsection{Required FogBus2 Components and their FL Functionality}\label{section:fl_implementation}

\begin{figure}[h]
    \centering
    \includegraphics[width=\linewidth,height=7cm]{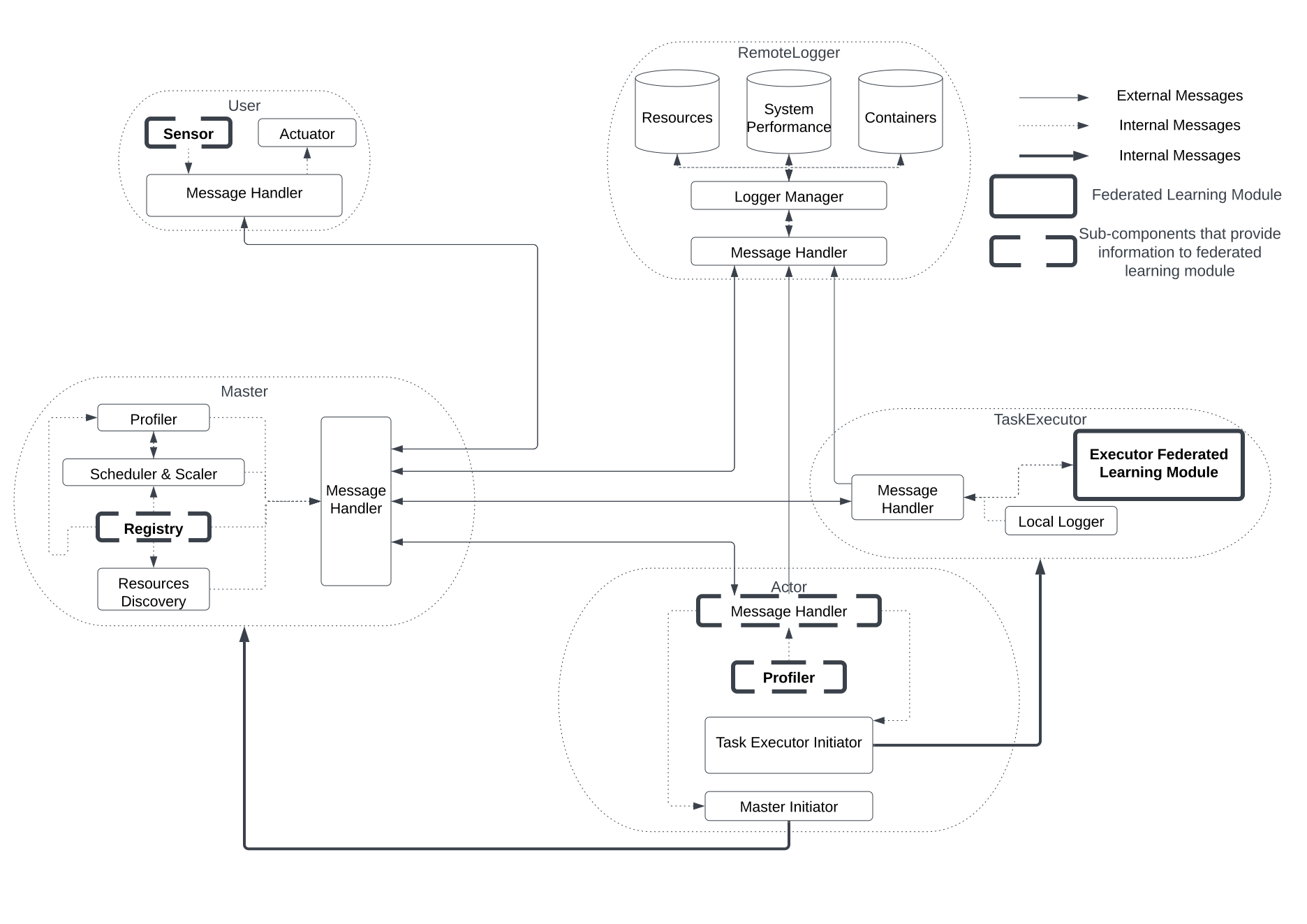}
    \caption{Federated Learning in FogBus2}
    \label{fig:3-1}
\end{figure}

The relationship between the FL module and FogBus2 framework is shown in Fig.~\ref{fig:3-1}. In this figure, bold dashed boxes show sub-components of FogBus2 that provide information to the FL module. Moreover, the FL module overrides the TaskExecutor component originally available in FogBus2. This section introduces how FogBus2 components can provide information to support the FL. Then, it describes how the FL module exists as tasks in the TaskExecutor component of FogBus2.
\subsubsection{Sensor}
The Sensor sub-component in the User component of FogBus2 is used to grasp input from users and forward it to the Master component, which will be further forwarded to the executor within the TaskExecutor component. Executors are classes that run FL tasks as functions, where function inputs are users' inputs stored in a dictionary. The Sensor sub-component is then configured to ask hyperparameters from users for the FL training process. While there can be various hyperparameters specific to different deployed FL algorithms, current hyperparameters collected from users include:
1) The model shared by participants cooperating in the FL training,
2) Total number of aggregations to perform on the aggregation server side,
3) Number of training epochs each worker has to complete before contributing the local model weights to the FL aggregation process,
4) Whether the FL training is conducted synchronously or asynchronously, and
5) The learning rate initially used by different workers to update the model.
When the FL application runs in FogBus2 and all corresponding TaskExecutors are ready, the Sensor sub-component collects those hyperparameters from users. It is the first step of the FL training. Next, different executors will conduct FL training based on hyperparameters.
\subsubsection{Registry}
In the FL implementation, different participants must regularly communicate with each other to transmit messages and model weights. This requires that different TaskExecutors running FL tasks know the IP address and port number of other FL TaskExecutors before FL training starts. In FogBus2, different tasks within an application are linked to each other according to a dependency graph. In the dependency graph, a task may be the parent of one/several tasks and be the child of one/several tasks. Also, results from parents will be forwarded to their children as input to task function calls on children. This feature is used to transmit the IP address and port number between different TaskExecutors of FL. The aggregation server is responsible for starting the FL training process by creating the FL model and calling selected workers to start training. This makes the Executor running the aggregation server know the network address of all other Executors running as workers so that it can send instructions. Other Executors running as workers only need to wait until the aggregation server contacts them, so the network address of the aggregation server is available, and they can send messages back. Thus, the implementation lets the task that operates as the aggregation server be a child task of all other tasks which run as workers. Moreover, the implementation lets the returning results of tasks that run as workers be the network address to which they are listening. These results will be inputs of the task running the aggregation server. In this way, the aggregation server can have the network address of all other workers before FL training starts.
\subsubsection{Message Handler}
Since Executors running as workers need to return the network address to which they are listening, they need to listen to that address. In order to make the Executor aware of the IP address for tasks running FL training, the implementation takes the IP address from the Actor module. The Actor component is responsible for starting the docker container of the TaskExecutor component in place. Thus, the Actor is physically on the same machine as the TaskExecutor, which makes the IP address of the Message Handler sub-component of the Actor module the same as the IP address of the Executor. The implementation adds a tag noting whether input data from users are related to FL or not. When the Actor component calls the TaskExecutor component to start tasks, it will add the IP address of its Message Handler to the input dictionary if the tag indicating tasks are related to FL. Since the Executor running FL tasks needs to communicate with each other regularly, the implementation lets them communicate with a separate port instead of the port used by the Message Handler sub-component to avoid congestion. The port is subject to the port availability of the machine running TaskExecutor. In this way, the TaskExecutor can start the socket server listening on the same IP address and different port compared to the Message Handler of the Actor, initializing it for FL purposes.
\subsubsection{Profiler}
The Profiler sub-component in FogBus2 is responsible for collecting statistics related to available system resources. FL optimization depends on parameters describing available system resources. Since the Actor initializes the TaskExecutor, the Profiler sub-component within the Actor is physically on the same machine as the Executor sub-component within TaskExecutor. Thus, data from Profiler of Actor also describe available resources for the Executor running FL tasks. The aggregation server is responsible for making optimization decisions, so it requires system parameters from all workers. The FLight implementation uses the property that parent tasks will pass results to child tasks to pass profiling information from worker Executors to the aggregation Executor. This is achieved by adding profiling data to Executor running as worker results if tasks are tagged as relating to FL. The TaskExecutor that runs the aggregation server will then receive the profiling information, which can be exploited to implement different FL optimization mechanisms accordingly.
\subsubsection{Federated Learning Modules (Executor)}
After properly initializing the aggregation server Executor and multiple worker Executors, the necessary information is ready for FL tasks. In particular, worker Executors will start the socket server listening for instructions from the aggregation server. At the same time, the aggregation server Executor will own the network addresses of socket servers on workers, as well as profiling information describing available computing resources. Then, the aggregation server task call functions from the FL component to define the FL process and execute the training accordingly. It is worth noting that Executors running as workers still are kept alive by holding the socket server on a separate thread after returning the socket server address. This is different from other tasks in FogBus2 that will terminate after returning results for children's tasks. This is because worker tasks still need to regularly listen to instructions from the aggregation server to perform FL training.
\subsection{FL Main Sub-components}
\begin{figure}[h]
    \centering
    \includegraphics[width=\linewidth,height=7cm]{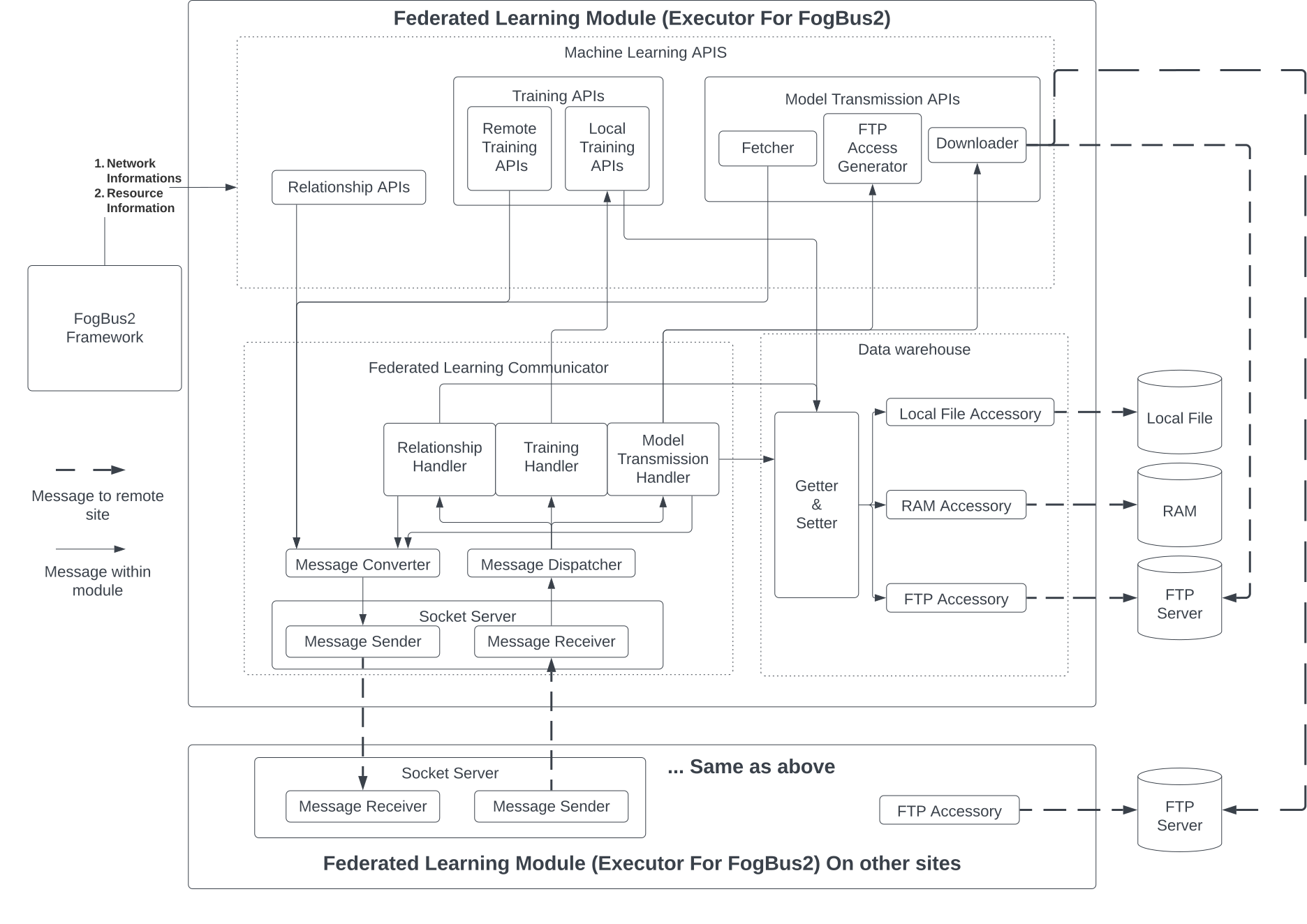}
    \caption{Federated Learning Component Structure}
    \label{fig:3-2}
\end{figure}

In this section, we describe how FogBus2's TaskExecutor component is extended to enable FL. Fig.~ \ref{fig:3-2} shows the design of FL implementation and its sub-components.
\par
The FL components take the network address of other Executors and statistics of available resources from Executors of the FogBus2 framework. The FL component is divided into three sub-components: 1) ML APIs, 2) FL Communicator, and 3) Data warehouse. The Data warehouse sub-component allows easy storage and data access required by FL. Moreover, the FL Communicator is used for communicating messages between different participants of FL training and handling them correspondingly. Lastly, ML APIs are sub-components that encapsulate FL logic such that calling those APIs is sufficient to define FL training. Overriding those APIs enables new ML models to be trained by FL.
\subsubsection{Data Warehouse Sub-component}
\label{subsection:dw}
\begin{figure}[h]
    \centering
    \includegraphics[width=\linewidth,height=5cm]{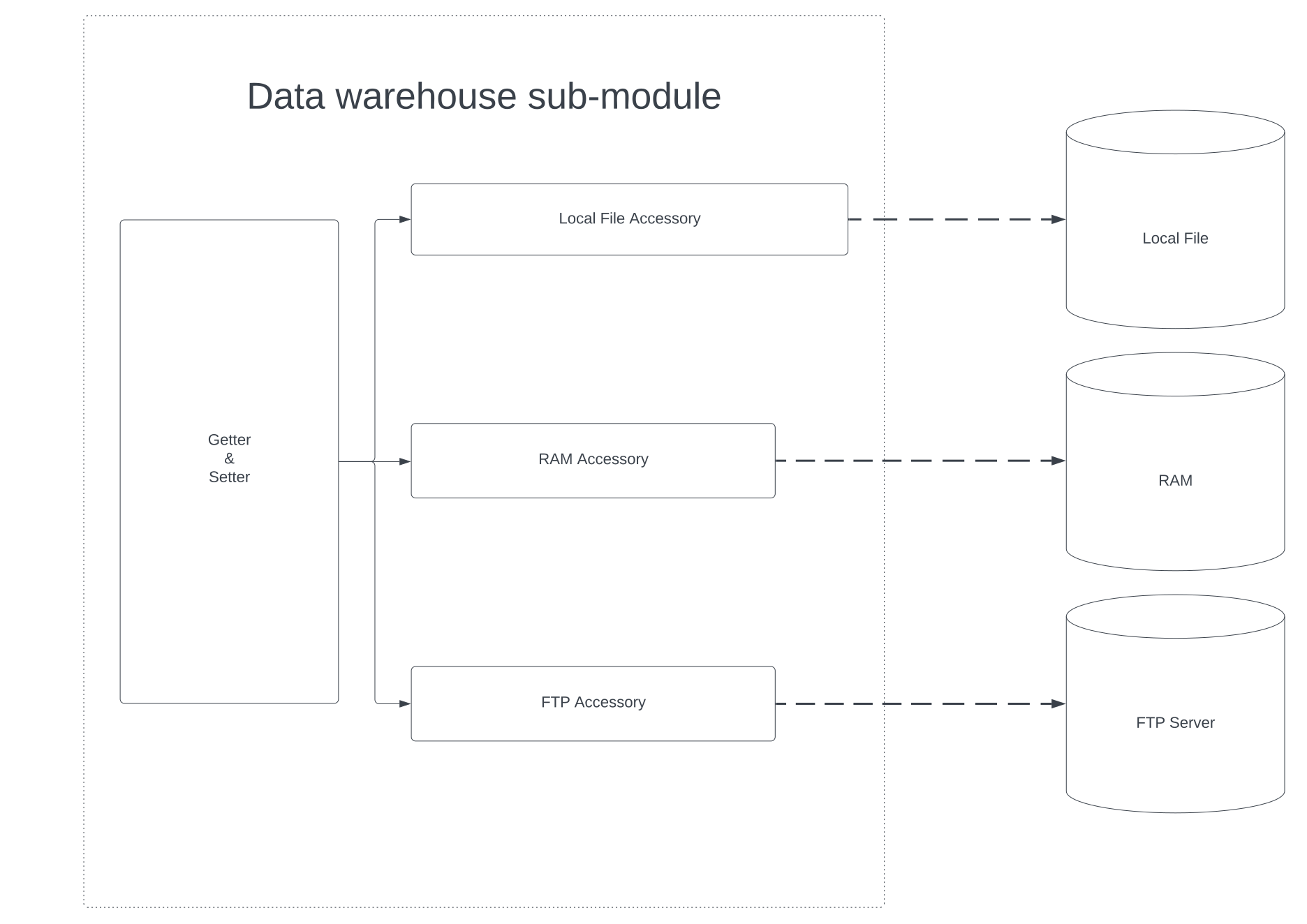}
    \caption{Data warehouse Design}
    \label{fig:3-3}
\end{figure}
\par
Figure~\ref{fig:3-3} shows the design of the data warehouse sub-component. This sub-component is responsible for providing an interface to access and store all kinds of data related to FL training. In FL training, data that needs to be stored includes 1) ML classes, 2) Parameter weights of ML models, 3) Parameter weights of ML models of other participants, and 4) Training Data. These four types of data can be placed in different storage, including RAM, remote repository, database, or files on local storage. However, to write and retrieve data from these storages, different implementations are required. The data warehouse sub-component encapsulates these various implementations. The data warehouse provides the getter and setter functions such that all kinds of data can be accessed by providing a unique ID. Moreover, if data is saved to the warehouse for the first time, the sub-component will return an ID that uniquely identifies that data. When a unique ID is provided to the getter function, the data warehouse uses the ID to retrieve the saved credentials and storage type used to store the data corresponding to the provided ID. Then it will use those credentials to retrieve the actual data. While setters allow data to be stored on a specified type of storage, default storage for ML model weights and training data is set to the local disk.
\par
This design allows extension for different storage types. Defining a new storage type can be done by defining the methods to write and retrieve data and then adding that to the data warehouse. Due to the design that only a unique ID is sufficient to retrieve data, the ML model, which is one type of data, can be referred to only by the ID locally. Referring to a remote ML model needs to specify the network address as well. This gives rise to the idea of the Pointer class, used by FL training participants to identify a model on a remote site uniquely. The Pointer class consists of the data warehouse's network address and unique ID for it. For example, when the aggregation server asks a remote worker to conduct training, multiple worker network addresses can be saved, and each worker can own several ML models. The aggregation server can provide the address to uniquely identify the worker and the unique ID to identify the model on that worker.
\subsubsection{Communication Sub-component}
\begin{figure}[h]
    \centering
    \includegraphics[width=\linewidth,height=4.4cm]{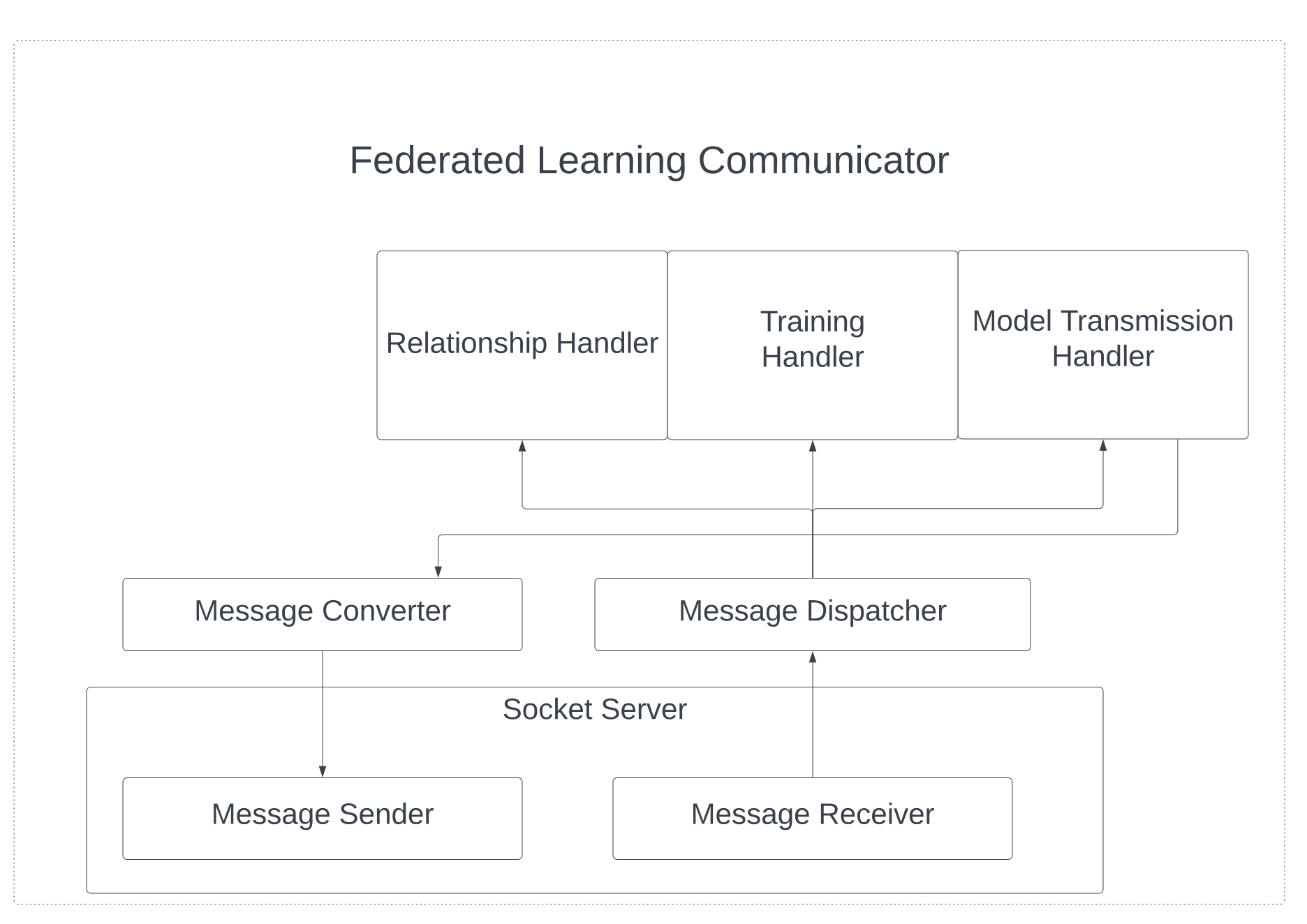}
    \caption{Federated Learning Communicator}
    \label{fig:3-4}
\end{figure}
Fig.~\ref{fig:3-4} shows the design of the FL communicator. Since FL training involves frequent message communication between different parties, the FL implementation provides its communicator. The FL communicator's sub-component consists of a socket server, a message converter, a message dispatcher, and handlers. The socket server is a server that listens to incoming messages via a receiver and sends out messages by the sender. All messages that go through the socket server are binary data. The message converter is responsible for converting messages into binary formats so that data can be transmitted between the local message sender and remote message receiver. The message dispatcher uses message types to forward messages to corresponding handlers. There are three handlers in the implementation. Firstly, the relationship handler is responsible for handling incoming requests for establishing an FL relationship. For example, if the aggregation server asks a remote site to be the worker of itself, then the relationship handler on that remote site is responsible for handling related messages. Secondly, training handlers are responsible for handling messages related to FL training. This includes the aggregation server asking a worker to start training, and the worker acknowledges to the server that local training is complete. Lastly, model transmission handlers are responsible for sending a request to fetch the weights of a remote model and send back the credentials required to download the weights.
\par
\begin{figure}[t]
    \centering
    \includegraphics[width=\linewidth,height=3.5cm]{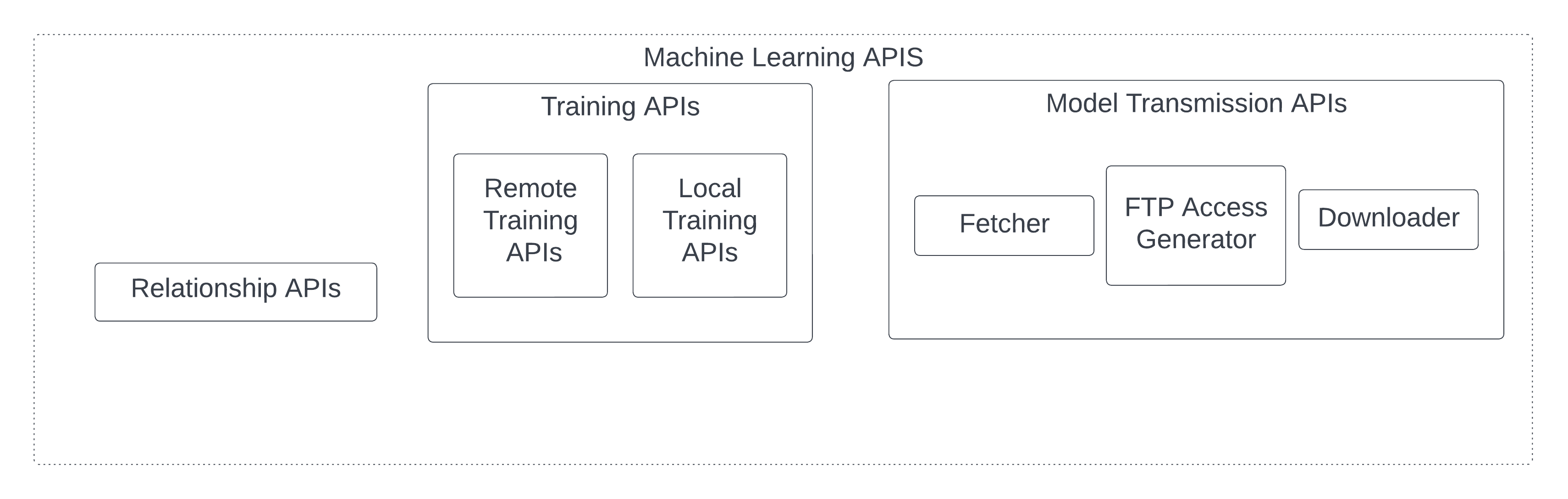}
    \caption{Machine Learning APIs}
    \label{fig:3-5}
\end{figure}

It is important to note that the weights are not transmitted directly through the socket connection between the message sender and receiver. This is because model weights are large compared to other messages. If weights are sent over the communication channel of FL, other messages have to wait for the weights to finish transmission. This long waiting time influences the time efficiency if the weights are sent over the communication channel of FL. Alternatively, in FLight, when a server receives the request to fetch a local model weight, the server saves the weights to a File Transmission Protocol (FTP) server and sends back a one-time login credential. The remote server fetching the model weights can use the credential to log in to the FTP server and download through FTP.
Fig.~\ref{fig:3-5} shows the design of the ML APIs sub-component. This sub-component is a minimum set of functions that an ML model requires to define for the training of different participants. All these functions are collected into a single class. Therefore, ML models can override functions to be deployed for FL training. ML APIs are further divided into relationships, training, and transmission APIs.
\subsubsection{ML API Sub-component}
\par
Relationship APIs are functions to establish a relationship with other models. It includes functions to request remote servers to act as its workers or request being workers of a remote aggregation server. Functions related to relationships first send a message via the communication sub-component, and the relationship handler on other sites will handle the forwarded message. Training APIs are functions related to ML training, including remote and local training functions. Remote training APIs are functions used by aggregation servers to request a remote worker to perform specified rounds of training. Those functions need a pointer class that refers to a remote model as an argument. So the message sender within the FL communication sub-component can use the network address within the pointer to send the message to a remote message receiver, and the handler can use a unique ID within the pointer on the remote side to retrieve the model via getter of the data warehouse module. Local training APIs are functions used to conduct some calculations on a locally stored model. These include functions to conduct training based on available data, which is the same as regular ML training. Moreover, local training APIs include various functions to federate model weights from workers for the aggregation server based on different algorithms. Besides, Model transmission APIs are functions used to transfer model weights between different participants. Fetcher functions are responsible for sending messages requesting remote model weights. After receiving the fetch request, the FTP access generator on the remote side generates a credential that the participant uses to fetch the model and downloads the model weights from the FTP server. The actual download functionality is implemented in the Downloader.
\subsection{FLight's Sub-components Interactions}
The ML APIs, FL communicator, and data warehouse sub-components enable FL mechanisms. This section presents how these different sub-components cooperate together to enable important functionalities, including the addition of a worker, model transmission between different servers, and conducting the training. 
\subsubsection{Worker Addition}
\begin{figure}[h]
    \centering
    \includegraphics[width=\linewidth,height=6cm]{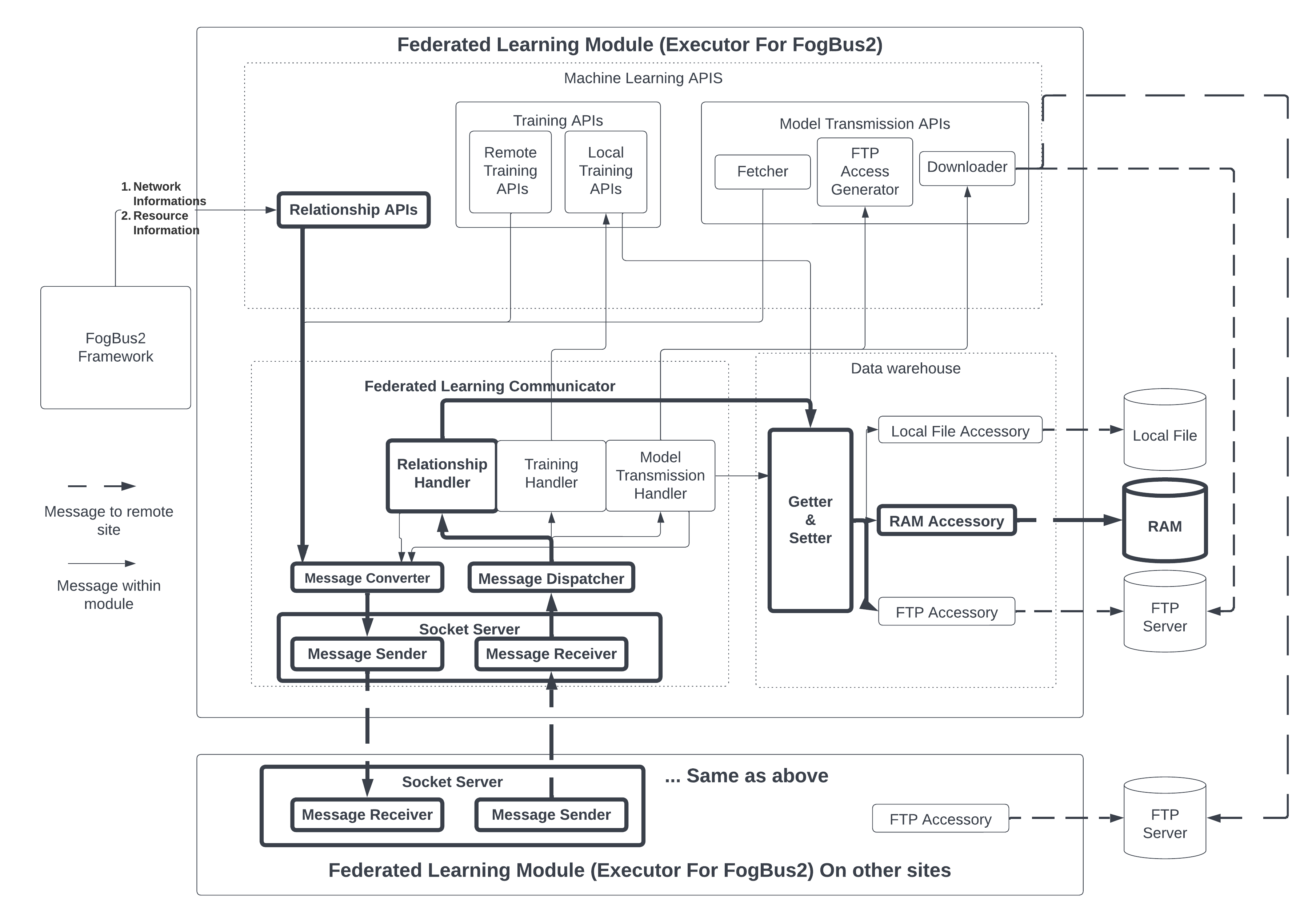}
    \caption{sub-components involved in worker addition}
    \label{fig:3-6-1}
\end{figure}

\begin{figure}[h]
    \centering
    \includegraphics[width=\linewidth,height=6cm]{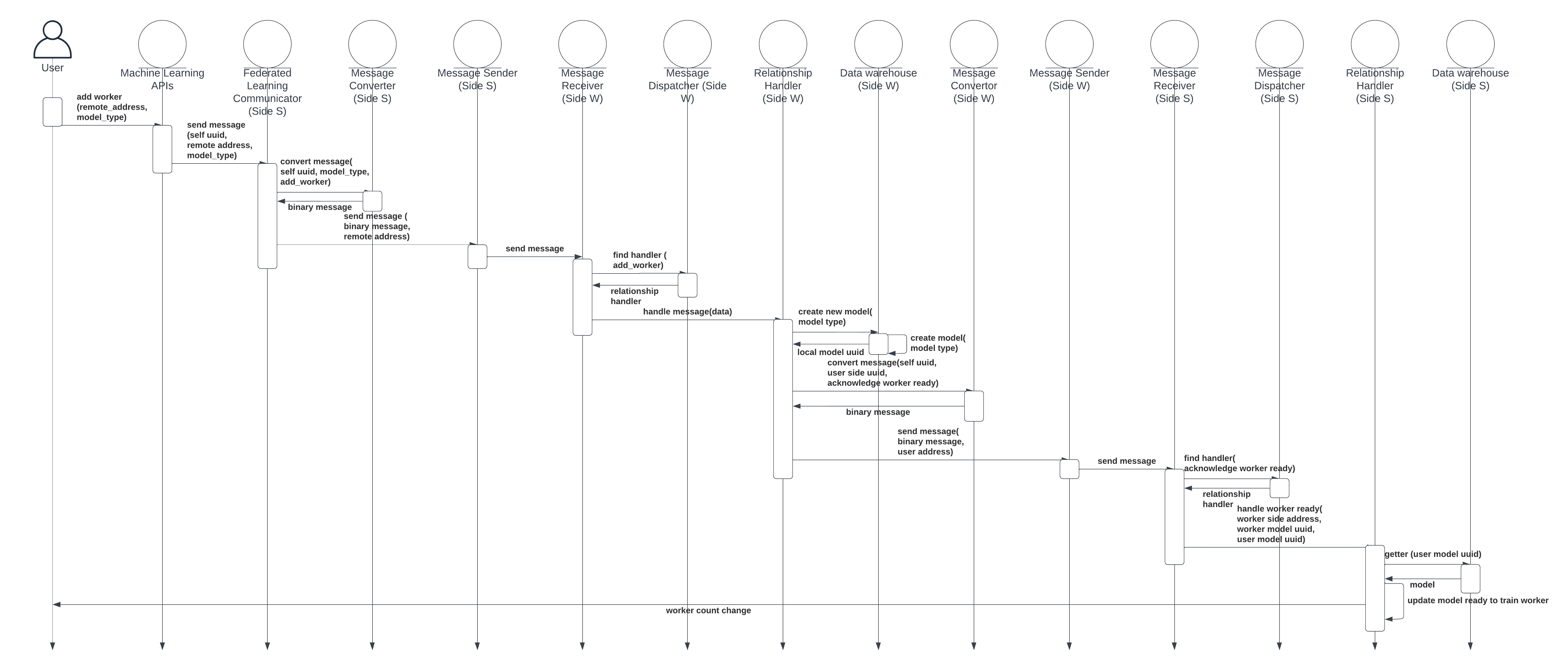}
    \caption{Adding worker sequence diagram}
    \label{fig:3-6-2}
\end{figure}
The sub-components to add a remote server as a worker is shown in Fig.~\ref{fig:3-6-1}, while Fig.~\ref{fig:3-6-2} depicts the corresponding sequence diagram. The aggregation server is called side $S$, and the worker is denoted by $W$. The function is called on side $S$ initially, which contains the following steps: 1) Before calling the function, an ML model should be created on the side $S$. 2) User first calls the relationship function in the ML APIs sub-component for worker addition on $S$. Then, $S$ requests $W$ to create an ML model with the same structure and initiate the worker model. 3) After calling the function of worker addition, the message sender on $S$ is called to send an invitation to the remote participant regarding the network address. Function arguments also include the unique ID of the aggregation server model. Then, $W$ creates a pointer class composed of the network address of the message sender on side $S$ and a unique ID of the aggregation server model. The remote worker model can use this pointer to refer to its server model. 4) The communicator calls the message converter on $S$ to pack the message into a tuple and serialize it into binary data for socket transmission. 5) After receiving data in binary format, the message sender on $S$ sends the data over the socket to the message receiver on $W$. 6) The message receiver on $W$ interprets the message and sends the remaining messages to the relationship handler on $W$. 7) The relationship handler on $W$ creates a model with an identical structure to the aggregation server model on $W$. The model is also added to the data warehouse sub-component for later retrieval. 8) The worker model on side $W$ saves the unique ID of the aggregation server model and the $S$ network address as the server model's pointer. Next, the worker model is ready for further instructions, such as training. 9) The relationship handler on $W$ informs $S$ that the worker model is ready. The message contains the unique ID of the worker model and the aggregation server model, as well as the network address of the $S$. Then the message passes through the converter, which serializes the message and transmits the message through a socket. 10) After $S$ receives the acknowledgment from $W$ that the worker model is ready, it lets the relationship handler on $S$ handle the message. 11) The relationship handler on $S$ uses the server ID to retrieve the aggregation server model from the data warehouse. After that, the pointer referring to the worker model, which consists of the unique ID of the worker model and the network address of $W$, will be recorded into the aggregation server model class. After these steps, the aggregation server model on $S$ has one extra stored worker model pointer. Moreover, the worker model on $W$ has one stored server pointer, referring to the aggregation server model.
\subsubsection{Transfer a model}
\label{subsection:comm_model}
\begin{figure}[h]
    \centering
    \includegraphics[width=\linewidth,height=6cm]{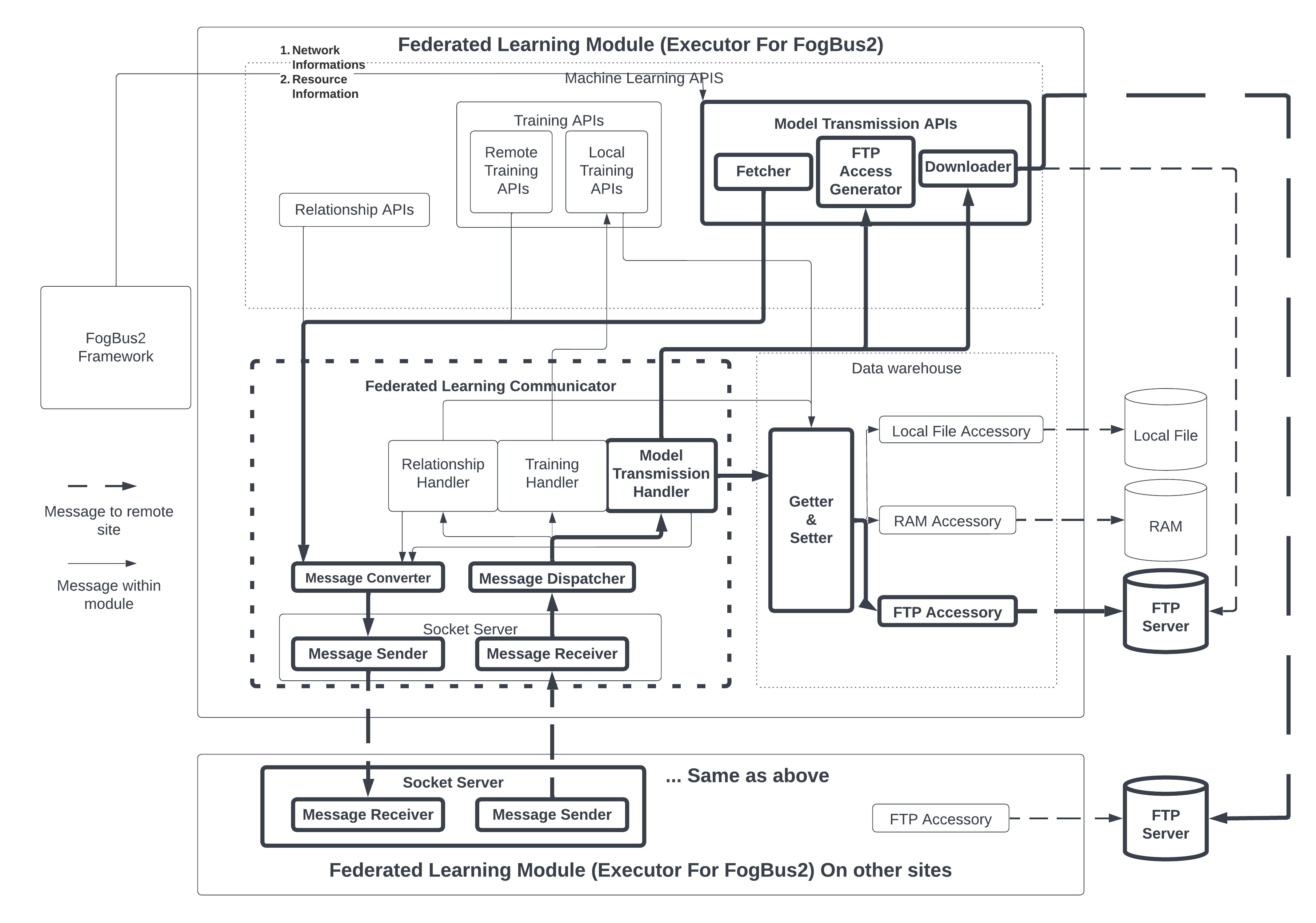}
    \caption{Components for Transferring model weights}
    \label{fig:3-7-1}
\end{figure}
\begin{figure}[h]
    \centering
    \includegraphics[width=\linewidth,height=6cm]{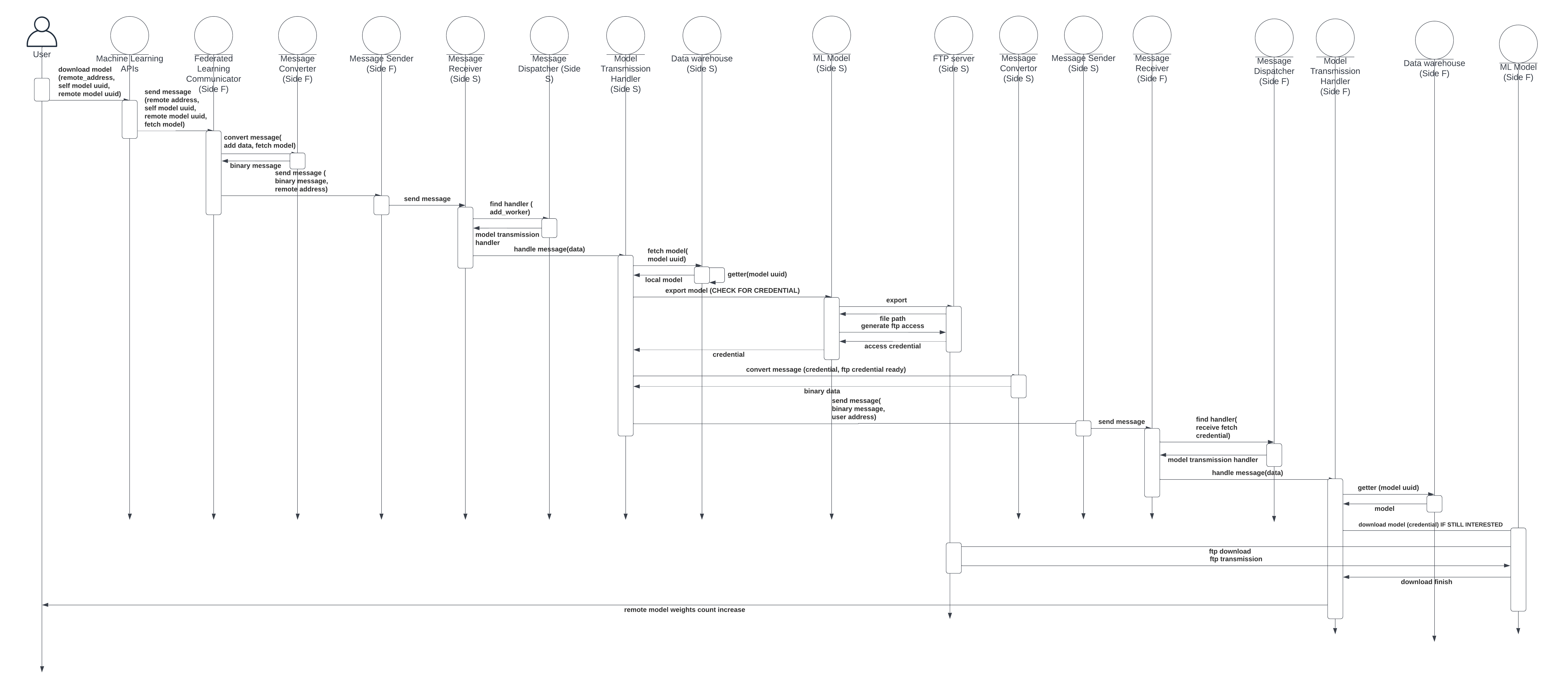}
    \caption{Model transmission sequence diagram}
    \label{fig:3-7-2}
\end{figure}
Fig.~\ref{fig:3-7-1} shows the sub-components involved in communicating model weights between servers, while Fig.~\ref{fig:3-7-2} depicts the respective sequence diagram. We assume the server fetching the model weight is $F$, and the server that sends back model weights is on side $S$. 1) A model on $F$ calls a fetching model function within the model transmission APIs. Arguments include a pointer referring to the remote model from which the local model fetches weights. Moreover, identification information, such as the unique ID of the model on $F$, is also provided. 2) Message is then serialized by a message converter and sent out by the message sender on $F$. The message converter will also add an additional tag to indicate that the message is about fetching the model so that the remote handler can react correspondingly. 3) When the message receiver on $S$ gets the message, the dispatcher forwards the message to its own model transmission handlers. The handler checks the pointer that $F$ has sent and uses the ID to retrieve the ML model from the data warehouse. 4) If the model exists, $S$ also checks whether it has the privilege to access its weights. Since model weights are not shared in public, it has to check for the identity of remote servers fetching it. 5) If the access check passes, the model transmission handler exports the model weights to a file in the FTP server. 6) The model transmission handler on $S$ collects the file name where model weights are stored and also login credentials for downloading that file from the FTP server as a response. 7) The credential will be sent back from $S$ to $F$. 8) After the message receiver on $F$ gets the message and forwards it to the model transmission handler, the handler sends out a fetch request. 9) If the check suggests the local model still wants remote model weights, then the model transmission handler will use the Downloader function to log in to the FTP server and download model weights to a local file. After these steps, a file containing the latest model weights of a remote model at the time of fetching is generated on $S$.

\subsubsection{Requesting to train a model}
\begin{figure}[h]
    \centering
    \includegraphics[width=\linewidth,height=6cm]{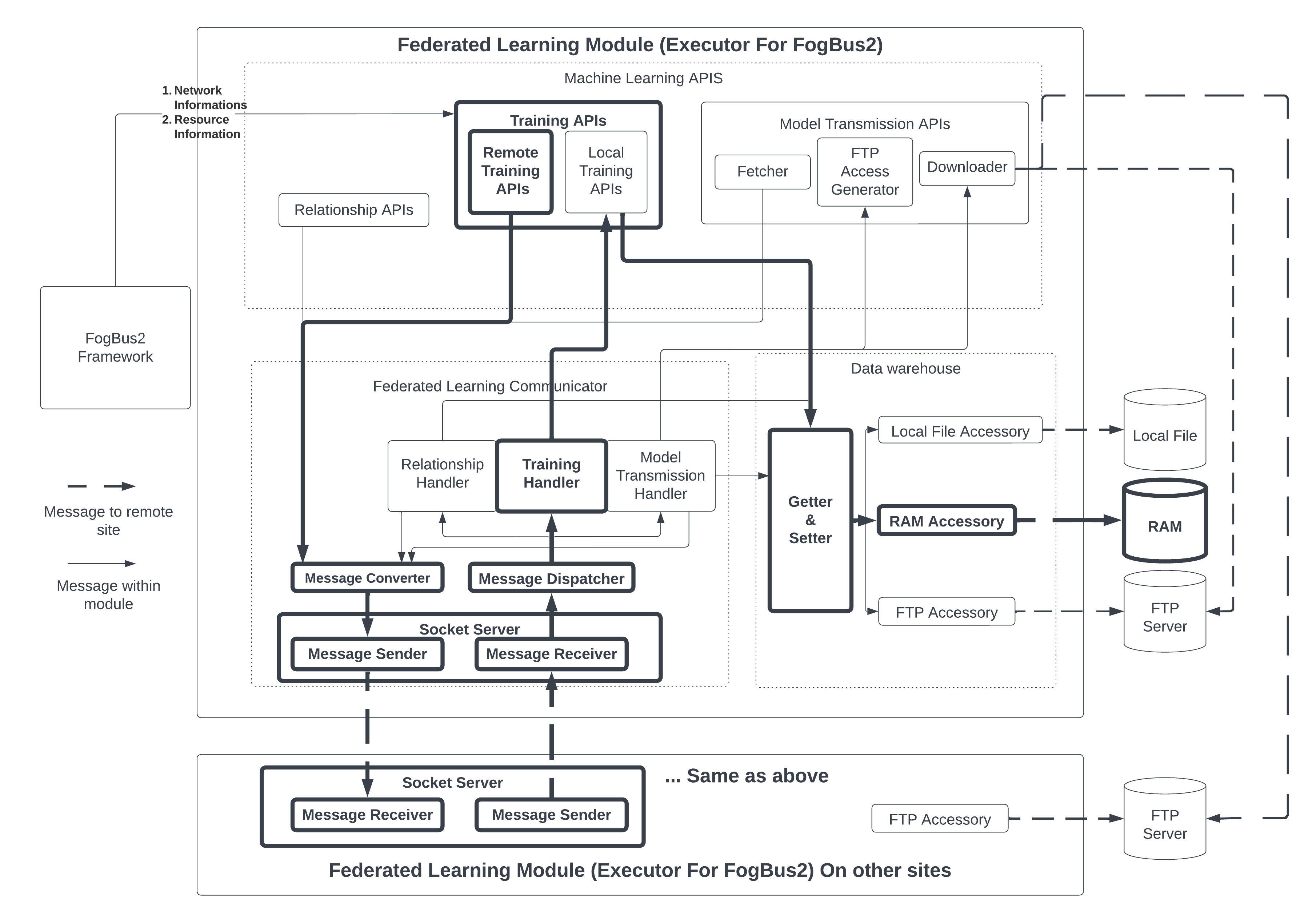}
    \caption{Components for worker training}
    \label{fig:3-8-1}
\end{figure}

\begin{figure}[h]
    \centering
    \includegraphics[width=\linewidth,height=6cm]{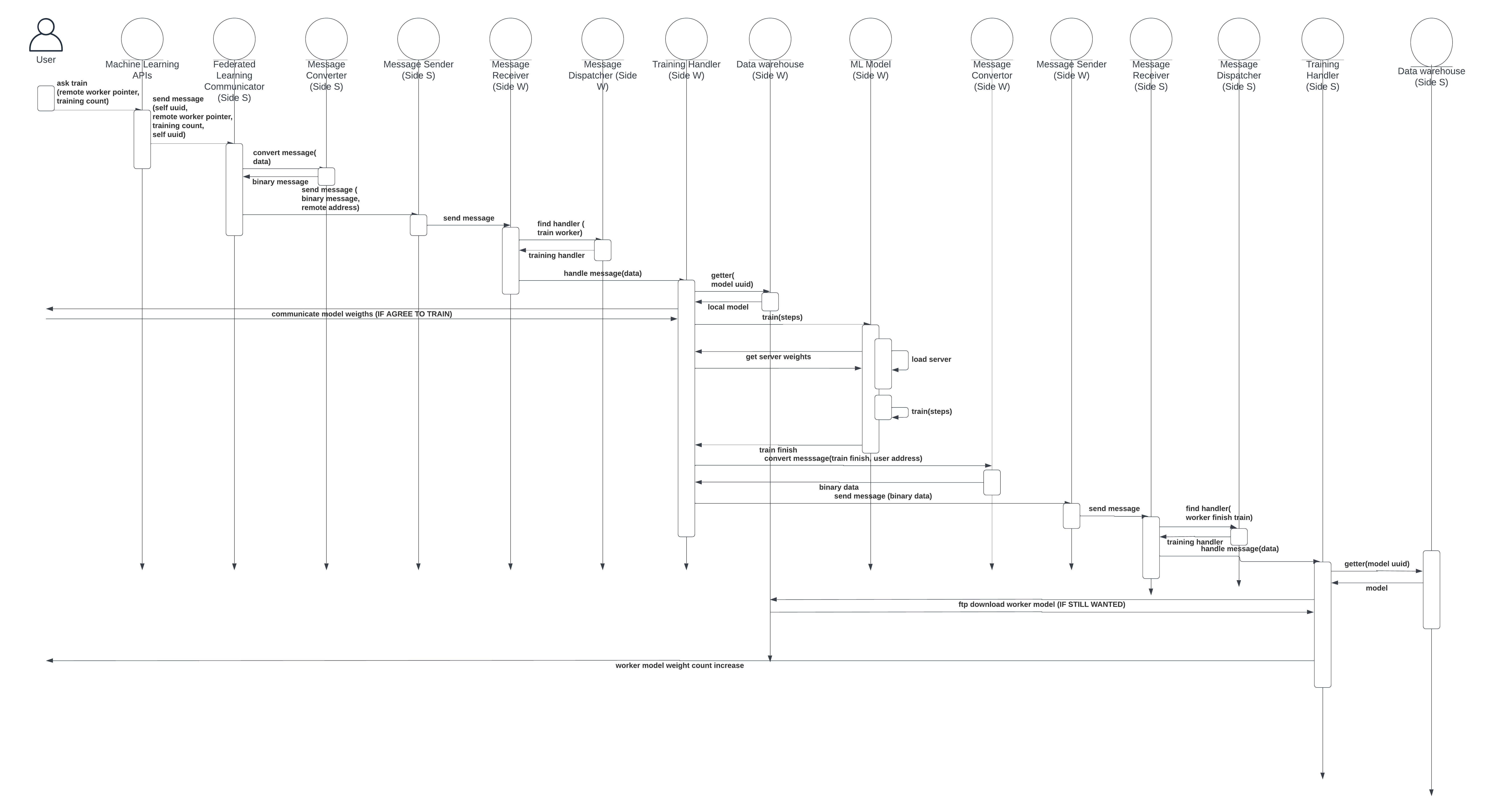}
    \caption{Remote worker train sequence diagram}
    \label{fig:3-8-2}
\end{figure}

Fig.~\ref{fig:3-8-1} shows the sub-component involved when an aggregation server asks a remote worker to train, while Fig.~\ref{fig:3-8-2} depicts the respective sequence diagram. Let's represent the aggregation server by $S$ and the remote worker by $W$: 1) Starting from $S$, the user calls the function from the Training APIs sub-component in an aggregation server model. The argument points to a remote worker model asking to train and the number of training epochs. 2) The function call in APIs serializes the message by message converter on $S$. The serialized message is then forwarded to site $W$ with a specified network address. 3) After receiving the message asking the worker model to conduct training, the dispatcher on $W$ returns the training handler to handle the message. 4) The training handler first retrieves the local worker model based on the pointer that $S$ provided. Next, the training handler checks whether the local model agrees to conduct specified training or not. Reasons for rejecting training instructions from a remote server may include the remote server is not recognized by the worker model or there are insufficient computation resources at the moment. 5) The worker model on $W$ fetches the aggregation server model weights. 6) After receiving the aggregation server model weights from $S$, the original parameter weights of the worker model will be replaced by aggregation server model weights. Next, the model is trained based on a specified epoch number. The training data comes from local data files, and the FLight framework support reading from the database for real-time applications or a local file for testing purposes. 7) After the training, $W$ acknowledges the training is done by sending the message back to $S$. The acknowledgment message contains pointers to the worker model and the aggregation server. 8) When the acknowledgment is received on $S$, the training handler controls the remaining message. It retrieves the aggregation server model from the data warehouse module and then checks if the aggregation server model still requires results from $W$. This is because the aggregation server can finish multiple rounds of aggregation while the worker model on $W$ conducts training. In this way, model weights from $W$ are outdated and useless. In the asynchronous FL case, the aggregation server takes the results no matter how many rounds of aggregation have already been conducted. The criteria for whether accepting results from a worker can be overridden by other logic for other use cases. 9) If the model accepts the model weights from the worker model on $W$, it fetches the model weights. After these steps, the aggregation server has a new file containing the model weights of a remote worker model. The aggregation server model can use weights to update local weights via aggregation.
\subsubsection{Aggregating worker model weights}
When the aggregation server model receives enough model weights or reaches a time limit, it then aggregates the received model weights. The default implementation of the FLight for synchronous FL waits until a specified amount of model weights are downloaded from workers. The default implementation of FLight for asynchronous FL starts aggregation once it receives any model weights from any worker. Noting that this step does not involve any communication with remote workers since all model weights have been downloaded beforehand. Moreover, during the aggregation process, if some workers respond with their updated model weights, the aggregation server ignores them or keeps those model weights for the next round of aggregation rather than the current aggregation round.
\subsection{Worker Selection Algorithm Design}\label{section:ws}
This section describes two heuristic algorithms to select workers participating in FL. The worker selection algorithms can be applied to synchronous and asynchronous FL. Both heuristic algorithms depend on the required time to complete training an entire batch of data for one epoch ${T_{one}}$ and the time required to transmit the model ${T_{transmit}}$; however, they differ from the perspective of acceptable maximum training time. 
\subsubsection{R-min R-max based worker selection}
Algorithm \ref{algo:rminmax} demonstrates the first heuristic algorithm.
\begin{algorithm}
\caption{R-min R-max based algorithm}\label{algo:rminmax}
\hspace{2pt} \textbf{Input}: 
\\ $W$: A set of workers
\\ $T_{one_w} \forall w \in W$: Training time required to go through all training data for one epoch
\\ $T_{transmit_w} \forall w \in W$: Time required to communicate model weights
\\ $rmin$: Minimum training epoch
\\ $rmax$: Maximum training epoch
\\
\hspace{2pt} \textbf{Output}: $W_{selected} \in W$
\\ $T_{min_w} \gets T_{one_w} * rmin + T_{transmit_w} \forall w \in W$
\\ $T_{max_w} \gets T_{one_w} * rmax + T_{transmit_w} \forall w \in W$
\\ $T_{minimum} \gets \min{T_{max_w}} \forall w \in W$
\\ $W_{selected} \gets \forall w \in W: T_{min_w} >= T_{minimum}$
\end{algorithm}
\par
After each round of aggregation, $rmin$ and $rmax$ are updated based on average accuracy among all selected workers. Let $accn_n$ be the accuracy achieved at the aggregation server on round n, and $accn_{n-1}$ be the accuracy achieved at the last round. Then $rmin$ and $rmax$ are updated:
\begin{equation}\label{equation:rmin}
rmin \gets rmin * \frac{accn_n+1}{accn_{n-1}+1}
\end{equation}

\begin{equation}
\label{equation:rmax}
rmax \gets rmax *\frac{accn_{n-1}+1}{accn_{n}+1}
\end{equation}

Firstly, this algorithm takes the training time required for each worker to go through their training data for one epoch as input. Also, the time required for communicating the model weights between the aggregation server and a worker is considered. After that, $rmin$ and $rmax$ are two hyperparameters that define the minimum and the maximum number of epochs a worker should train before sending back the worker model weights to the aggregation server. 
\par
If a worker trains for an insufficient amount of epochs before responding to the aggregation server, then model weights from that worker will have limited differences compared to the original model weights obtained from the aggregation server. $rmin$ is then selected to let the worker contribute model parameter weights with a promising difference. Among different ML models and available data sizes, difference $rmin$ needs to be figured out to ensure a meaningful update from workers. $rmax$, on the other hand, defines the maximum number of epochs a worker can train before responding. If a worker trains for too many epochs before aggregation, the model weights will be biased to training data that the worker locally uses. This will result in a large divergence between the aggregation server's update trend and the worker. In order to keep workers updating model weights in a similar direction as the average direction among all workers, regular communication with the aggregation server is necessary. This makes the algorithm introduce $rmax$ to limit workers from training too many epochs locally. After selecting $rmax$ and $rmin$, the algorithm calculates the time required to train those amount of epochs plus transmission time for each worker. This time can also be interpreted as the required time after the aggregation server sends out the instruction to conduct training until the aggregation server receives a response. Although this time can be varied since estimation for $T_{one_w}$ and $T_{transmit_w}$ has a difference with reality, it provides a heuristic suggesting which worker will respond faster. The selection criteria intend to minimize the time that fast computing workers wait for slow computing workers. When there is a difference between the time required to finish specified training, fast computing workers can train for more epochs than slow computing workers, which allows them to respond to the aggregation server in a similar time. However, extra training rounds from fast computing workers cannot exceed $rmax$. For slow computing workers, the minimum requirement is complete $rmin$ rounds of training. This suggests the selection criteria described in algorithm \ref{algo:rminmax} (lines $3 - 4$). If a worker requires more time to train a minimum number of epochs compared to the worker that can finish the maximum number of epochs for training, then that worker is excluded. After excluding slow computing workers, it is guaranteed that within the time the fastest computing workers finish maximum epochs of training, all other selected workers can at least complete the minimum training requirement. In order to achieve time efficiency of training, the worker selection algorithm lets fast computing workers participate in earlier training rounds. The initial worker selection process guarantees that only fast computing workers are selected. In order to generally include slow computing workers as training proceeds, the update will decrease $rmin$ while increasing $rmax$. Increasing the upper limit and decreasing the lower limit has the following impact: 1) Since the maximum number of iterations a worker can train before aggregation increases, this also increases the time required to complete full training rounds for all workers. Consequently, it increases the minimum value among those times. 2) In the same way, with decreasing $rmin$, the minimum requirement for workers decreases, resulting in reducing the time required for each worker to complete minimum training requirements. 3) According to the selection criteria, a worker will be selected only when they can finish the minimum required training before the fastest worker completes the maximum allowed amount of training between aggregation on the server side. With decreasing time, slow-computing workers are required to finish minimum training, and with increasing time, fast-computing workers need to finish maximum training epochs. Slow-computing workers can be included since they can finish minimum training before fast-computing workers finish the maximum allowed training amounts. Consequently, decreasing $rmin$ while increasing $rmax$ as the training progress can provide the effect of fast computing workers joining in an earlier round and slow computing workers joining later, which is time efficient. Training progress is expressed as an increase in the accuracy achieved by aggregated model weights against testing data. According to Eq. \ref{equation:rmin} and Eq.~\ref{equation:rmax}, $rmin$ is updated by multiplying the accuracy of previous aggregation rounds and dividing it by the accuracy in the current round. So, the more significant increase between the accuracy achieved by the aggregation server model in two aggregation rounds, the faster $rmin$ drops and vice versa for $rmax$. Furthermore, these equations adjusts the numerator and denominator by adding one to avoid the situation in which ML model accuracy surges in earlier training rounds. Without the adjustment, when there is a significant increase in accuracy, the factor deciding the decrease in $rmin$ is going to be very large. This will cause $rmin$ to decrease very fast and hit a low value in earlier training rounds. The same condition applies to $rmax$ in terms of increase. If $rmin$ reaches a low value and $rmax$ reaches a very large value, then a large proportion of workers is eligible based on the selection criteria. This causes slow-computing workers to be included in training too early.
\paragraph{Discussion of R-min R-max worker selection}
The algorithm addressed above has the potential to accelerate the FL training efficiency. However, the design has defects that may fail under specific scenarios.
\par
The first scenario is due to improper initialization of $rmin$ and $rmax$. If $rmin$ is initialized too low, it requires minimum time to satisfy the minimum training epochs. At the worker selection stage before the first round of training, since all workers have the relatively low time required to satisfy minimum training epochs, many slow-computing workers will be included. Including a large number of inefficient workers is harmful to training efficiency. On the other hand, initializing $rmin$ to a large value will result in a large training time required to satisfy minimum training epochs. This excludes a large number of workers in the early stage. Although workers selected under large initial $rmin$ are fast responding, inadequate workers participating in early-stage training can cause a large time before model accuracy starts to increase. Since $rmin$ only starts to drop once accuracy rises, slow accuracy growth in the early stage delays the time when more workers are included. This negatively affects training efficiency. The same scenario happens when the initialization of $rmax$ is chosen inappropriately. For every available machine learning model structure and training data, optimal initialization of $rmin$ and $rmax$ can be derived by grid search. However, a closed-form solution cannot be derived before training. This makes the worker selection algorithm \ref{algo:rminmax} hard to be applied to large categories of models. Secondly, the value of $rmin$ and $rmax$ may drop and increase too fast in the early stage. For ML training from scratch, model weights are initialized randomly. This causes the initial accuracy of the ML model to be relatively low compared to accuracy after a few epochs of training during the early stage. Since a significant accuracy increase leads to low $rmin$ and high $rmax$, the large difference between $rmin$ and $rmax$ arises when the accuracy surge happens in an earlier round of ML training. As a result, many slow workers will be selected in early rounds, which is time inefficient. If the ML model uses a pre-trained model, it can bypass the scenario that accuracy differences are too significant in earlier training rounds. However, this limits the types of applicable ML models to those models working with a pre-trained model. Furthermore, the issue of $rmin$ and $rmax$ diverging too quickly gets worse when the FL is conducted asynchronously. This is because asynchronous FL can aggregate results more frequently, causing a more frequent update of $rmin$ and $rmax$ and leading to a large diverge. Consequently, algorithm \ref{algo:rminmax} has the issue of hard initialization of $rmin$ and $rmax$ such that they do not diverge too quickly. During training, if the accuracy increases unstably, the values of $rmin$ and $rmax$ will be extremely low and large. This will cause a large proportion of slow workers to be selected, reducing time efficiency. Changing the FL from synchronous to asynchronous further worsened the situation.
\subsubsection{Training-time-based asynchronous FL}
Algorithm \ref{algo:tbased} is a modified version of algorithm \ref{algo:rminmax} that addresses the aforementioned issues.
\begin{algorithm}
\caption{Training-time-based worker selection}\label{algo:tbased}
\hspace{2pt} \textbf{Input}: 
\\ $W$: A set of workers
\\ $T_{one_w} \forall w \in W$: Training time required to go through all training data for one epoch
\\ $T_{transmit_w} \forall w \in W$: Time required to communicate model weights
\\ $r$: Worker training iteration
\\ $T$: Time allowed for this round of training
\\
\hspace{2pt} \textbf{Output}: $W_{selected} \in W$
\\1) $T_{total_w} \gets T_{one_w} * r + T_{transmit_w} \forall w \in W$
\\2) $W_{selected} \gets \forall w \in W: T_{total} <= T$
\end{algorithm}
This algorithm selects workers based on the time required to complete a specified amount of training. Moreover, the algorithm will update $T$, which refers to the time allowed for each round of training. Let $W_{ns}$ be the group of workers not selected yet, $accn_n$ be the accuracy achieved at the current round of aggregation, and $accn_{n-1}$ be the accuracy achieved last round. Let $A$ be the accuracy improvement threshold such that $T$ will only increase when the accuracy boost between two rounds of aggregation is less than that threshold. The update can be expressed in Eq.~\ref{equation:tupdate}.
{
\small
\begin{equation}\label{equation:tupdate}
T \gets \min{T_{total_w} \forall w \in W_{ns}} \text{  if  } accn_n - accn_{n-1} < A
\end{equation}
}
The idea of the worker selection algorithm \ref{algo:tbased} is that each worker should perform unified epochs of training before responding to the aggregation server. This allows calculating the total time required to conduct training and communicate model weights back as $T_{total}$. After that, a threshold time is selected, which excludes slow-computing workers from training. As the training progresses, if the aggregation server model's accuracy stops increasing, more workers are included. This is achieved by increasing the time limit. For algorithm \ref{algo:tbased}, the only hyperparameter that needs to be initialized is $T$. The initialization is straightforward, in which $T$ can be set to zero at the start. In this case, no worker can be selected, causing no increase in the accuracy. Thus, the update mechanism in Eq.~\ref{equation:tupdate} for $T$ is triggered. This allows more workers to be eligible for FL training. Initializing $T$ to zero or a small value has little impact on time efficiency. This is because if accuracy fails to increase for one epoch due to an insufficient worker selected, $T$ will increase and allow a more significant amount of workers to participate in FL training. The worker selection algorithm \ref{algo:tbased} together with the update Eq.~ref{equation:tupdate} sacrifice a little training time to allow the appropriate amount of workers to be selected, which has the potential to boost overall efficiency. Secondly, bringing in slower workers only when accuracy stops increasing means slow-computing workers are selected only when a converged accuracy is achieved from faster-computing workers. This prevents slow computing workers from joining in the early stage and only allows those slow workers to train when it is necessary to include them to achieve the desired accuracy, which improves time efficiency. Thirdly, $rmin$ and $rmax$ also diverge fast if FL is conducted asynchronously since there is more frequent aggregation and hence more frequent update of $rmin$ and $rmax$. Each update increases the difference between $rmin$ and $rmax$. The divergence between $rmin$ and $rmax$ leads to slow-computing workers being included in early training epochs. Algorithm \ref{algo:tbased}, on the other hand, is compatible with asynchronous FL. This is because even when more frequent aggregation is conducted, as long as the model accuracy keeps increasing, update Eq.~\ref{equation:tupdate} will not be triggered, preventing slow-computing workers from joining the training process.
\subsubsection{Estimated required time for training}
Both worker selection algorithms depend on the time required to communicate model weights and conduct training for each worker. When each worker is selected and the response model weights to the aggregation server are sent, the actual time consumed for communication and training is updated. Initially, the value is estimated by a heuristic. The estimated training time is based on system parameters provided by the FogBus2 framework. For the required time to transmit model weights, the estimated time is obtained based on the randomly transmitted model weights from the aggregation server to each worker to obtain the required time for the transmission. For training time, $T_{one}$ is the required time to train one epoch based on CPU availability and the CPU frequency of all workers. The aggregation server conducts training over one piece of data and records the consumed time as well as the CPU frequency allocated to conduct training. Moreover, the amount of training data that each worker contain is collected when the worker acknowledges they are ready to train. Then, $T_{one_w}$ is estimated:
{
\small
\begin{equation}\label{equation:t_one_calculate}
T_{one} \gets \frac{T_{onedata}}{CPU^{freq}_{s}} * CPU^{freq}_{w} * CPU^{prop}_{w} * N_{w} \forall w \in W
\end{equation}
}
The Eq.~\ref{equation:t_one_calculate} first estimates the time required to train one piece of data on each worker based on the time to train one data on the aggregation server and the multiplier between CPU frequencies. After that, the estimated required time for the training of each worker is obtained by multiplying by the obtained value to the amount of data belonging to each worker.
\section{Performance Evaluation}
\label{performance_evaluation}
This section presents the system configuration and training dataset used for the performance evaluation along with the results obtained from  different experiments.  
\subsection{System Configurations and Training Dataset}
To perform experiments, we have used a Mac Book Pro with 8 ARM-based CPU cores and 16 GB of RAM and a Desktop computer with 8 Core Core i9 CPU and 32 GB of RAM to run four Virtual Machines (VMs). One VM runs the aggregation server model, while the rest of the VMs run worker models. Each VM is allocated 2 GB of RAM, while the CPU core number and base CPU frequency of all VMs are the same. Three VMs evenly distribute all worker models for different numbers of worker models participating in FL. We conduct an FL with only one worker model, which simulates sequential implementation. Afterward, we use FL with 10-worker models and 30-worker models, in which workers are evenly distributed into three different machines. A VM has 3-4 FL worker models when there are 10 models in the FL, or a VM has 10 worker models when there are 30 models in total. For communication, each VM is assigned a separate network address through which different FL and FogBus2 components can communicate. 
\par
We have used two different datasets in experiments that are commonly used in other FL research: 1) MINST \cite{deng2012mnist} and 2) CIFAR-10 \cite{CIFAR}. These datasets have sufficient data such that both sets have 60000 training data. The data is split and distributed to different workers, ensuring all workers have a sufficient amount of distinct training data. The amount of training data allocated to each worker model in each experimental configuration when there are 10 and 30 worker models are shown in table~\ref{table:ten_worker} and table~\ref{table:thirty_worker}, respectively.

\begin{table}[]
\centering
\caption{Batch of data each worker is allocated (10 worker)}\label{table:ten_worker}
\resizebox{1\linewidth}{!}{%
\begin{tabular}{|c|c|c|c|c|c|c|c|}
\hline
 Config& Data set & W1 & W2/W3 & W4 & W5/W6 & W7 & W8/W9/W10 \\ \hline
1 & MINST & 10 & 0 & 0 & 0 & 0 & 0 \\ \hline
2 & MINST & 1 & 1 & 1 & 1 & 1 & 1 \\ \hline
3 & MINST & 1 & 0 & 3 & 0 & 0 & 2 \\ \hline
4 & CIFAR & 100 & 0 & 0 & 0 & 0 & 0 \\ \hline
5 & CIFAR & 10 & 10 & 10 & 10 & 10 & 10 \\ \hline
6 & CIFAR & 10 & 0 & 30 & 0 & 0 & 20 \\ \hline
\end{tabular}
}
\end{table}

\begin{table}[]
\centering
\caption{Batch of data each worker is allocated (30 worker)}\label{table:thirty_worker}
\resizebox{1\linewidth}{!}{%
\begin{tabular}{|c|c|c|c|c|c|c|c|}
\hline
 Config& Data set & W1 & W2 - W10 & W11 & W12 - W20 & W21 & W22 - W30 \\ \hline
1 & MINST & 30 & 0 & 0 & 0 & 0 & 0 \\ \hline
2 & MINST & 1 & 1 & 1 & 1 & 1 & 1 \\ \hline
3 & MINST & 4 & 0 & 8 & 0 & 0 & 2 \\ \hline
4 & CIFAR & 300 & 0 & 0 & 0 & 0 & 0 \\ \hline
5 & CIFAR & 10 & 10 & 10 & 10 & 10 & 10 \\ \hline
6 & CIFAR & 40 & 0 & 80 & 0 & 0 & 20 \\ \hline
\end{tabular}
}
\end{table}

Configurations 1 and 4 in table~\ref{table:ten_worker} and table~\ref{table:thirty_worker} only allocate training batches of data to one worker model to simulate sequential training. However, other configurations distribute training data to different worker models. Configurations 2 and 5 indicate the situation where each worker model holds an even amount of training data, while configurations 3 and 6 denote the case that training data is unevenly distributed. The total amount of available data for training among all workers is the same for configurations 1-3 and 4-6. Different amount of training data leads to a different time to complete the training among workers. All the training is conducted asynchronously and synchronously for one hundred epochs. The long training epoch ensures sufficient time for the aggregation server model to achieve potential accuracy under available workers.
\subsection{Performance Results}

\begin{figure*}[!ht]
\begin{subfigure}{.5\textwidth}
  \centering
  \includegraphics[width=0.8\linewidth, height=4cm]{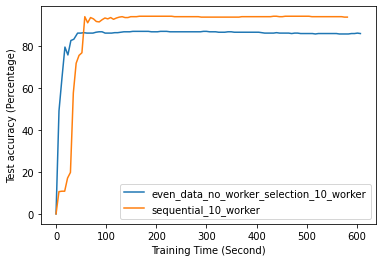}
  \caption{MINST 10 workers]}
\end{subfigure}%
\begin{subfigure}{.5\textwidth}
  \centering
  \includegraphics[width=0.8\linewidth, height=4cm]{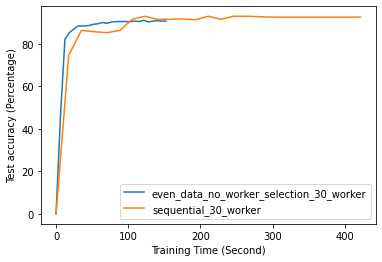}
  \caption{MINST 30 workers]}
\end{subfigure}
\begin{subfigure}{.5\textwidth}
  \centering
  \includegraphics[width=0.8\linewidth, height=4cm]{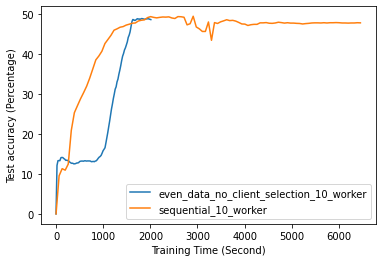}
  \caption{CIFAR 10 workers}
\end{subfigure}
\begin{subfigure}{.5\textwidth}
  \centering
  \includegraphics[width=0.82\linewidth, height=4cm]{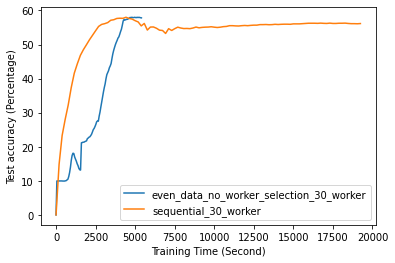}
  \caption{CIFAR 30 workers}
\end{subfigure}
\caption{Sequential Training VS FL Training (even data distribution, no worker selection)}
\label{fig:sequential_vs_even}
\end{figure*}
\begin{figure*}[!ht]
\begin{subfigure}{.5\textwidth}
  \centering
  \includegraphics[width=0.8\linewidth, height=4cm]{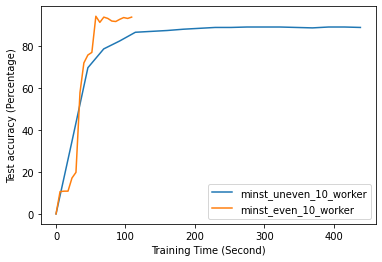}
  \caption{MINST 10 workers]}
\end{subfigure}%
\begin{subfigure}{.5\textwidth}
  \centering
  \includegraphics[width=0.8\linewidth, height=4cm]{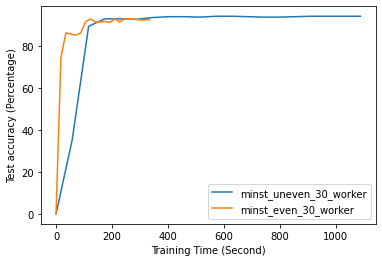}
  \caption{MINST 30 workers]}
\end{subfigure}
\begin{subfigure}{.5\textwidth}
  \centering
  \includegraphics[width=0.8\linewidth, height=4cm]{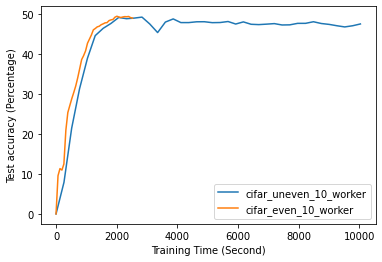}
  \caption{CIFAR 10 workers}
\end{subfigure}
\begin{subfigure}{.5\textwidth}
  \centering
  \includegraphics[width=0.8\linewidth, height=4cm]{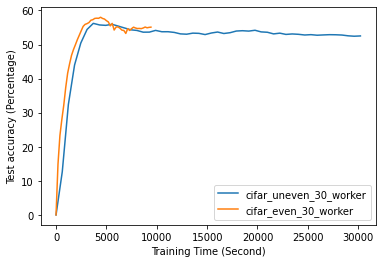}
  \caption{CIFAR 30 workers}
\end{subfigure}
\caption{Even vs. Uneven Data Distribution}
\label{fig:even_vs_uneven}
\end{figure*}
\begin{figure*}[!ht]
\begin{subfigure}{.5\textwidth}
  \centering
  \includegraphics[width=0.8\linewidth, height=4cm]{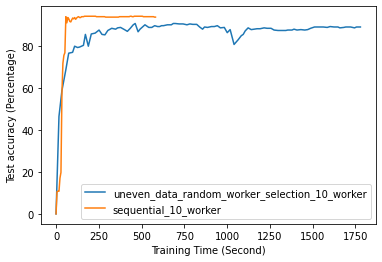}
  \caption{MINST 10 workers]}
\end{subfigure}%
\begin{subfigure}{.5\textwidth}
  \centering
  \includegraphics[width=0.8\linewidth, height=4cm]{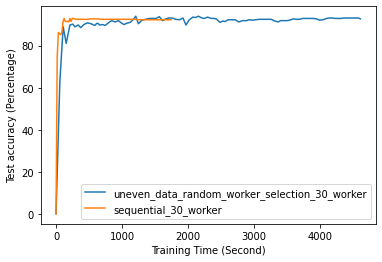}
  \caption{MINST 30 workers]}
\end{subfigure}
\begin{subfigure}{.5\textwidth}
  \centering
  \includegraphics[width=0.8\linewidth, height=4cm]{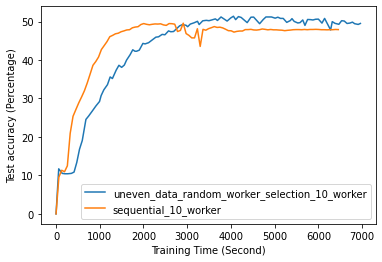}
  \caption{CIFAR 10 workers}
\end{subfigure}
\begin{subfigure}{.5\textwidth}
  \centering
  \includegraphics[width=0.8\linewidth, height=4cm]{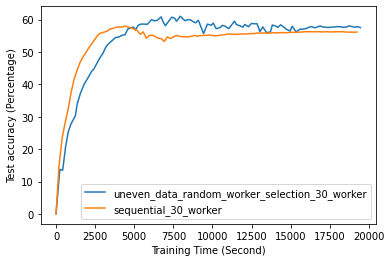}
  \caption{CIFAR 30 workers}
\end{subfigure}
\caption{Random worker selection vs. Sequential}
\label{fig:rws_vs_sequential}
\end{figure*}
\begin{figure*}[!ht]
\begin{subfigure}{.5\textwidth}
  \centering
  \includegraphics[width=0.8\linewidth, height=4cm]{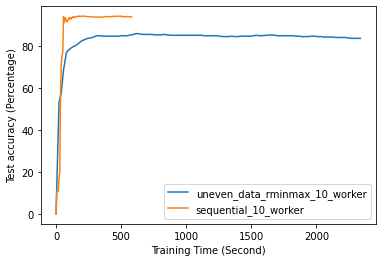}
  \caption{MINST 10 workers]}
\end{subfigure}%
\begin{subfigure}{.5\textwidth}
  \centering
  \includegraphics[width=0.8\linewidth, height=4cm]{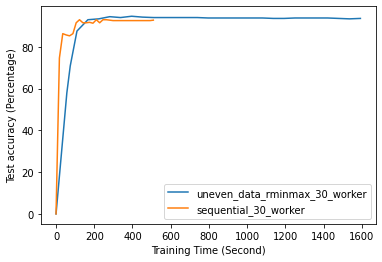}
  \caption{MINST 30 workers]}
\end{subfigure}
\begin{subfigure}{.5\textwidth}
  \centering
  \includegraphics[width=0.8\linewidth, height=4cm]{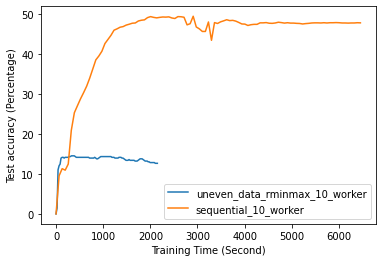}
  \caption{CIFAR 10 workers}
\end{subfigure}
\begin{subfigure}{.5\textwidth}
  \centering
  \includegraphics[width=0.8\linewidth, height=4cm]{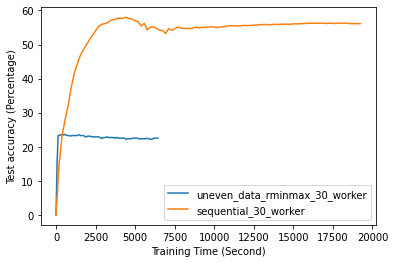}
  \caption{CIFAR 30 workers}
\end{subfigure}
\caption{R-min R-max Worker Selection vs. Sequential ($rmin$, $rmax$ initialise to 5)}
\label{fig:rminmax_vs_sequential}
\end{figure*}
\begin{figure}[!ht]
    \centering
    \includegraphics[width=0.8\linewidth,height=4cm]{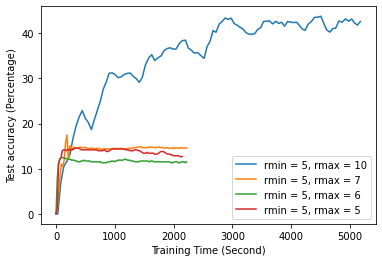}
    \caption{R-min R-max Worker Selection with Different R-max Initialisation}
    \label{fig:r-min-r-max-initialisation}
\end{figure}
\begin{figure*}[!ht]
\begin{subfigure}{.5\textwidth}
  \centering
  \includegraphics[width=0.8\linewidth, height=4cm]{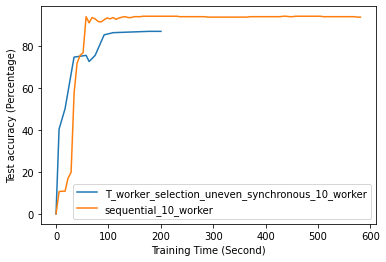}
  \caption{MINST 10 workers]}
\end{subfigure}%
\begin{subfigure}{.5\textwidth}
  \centering
  \includegraphics[width=0.8\linewidth, height=4cm]{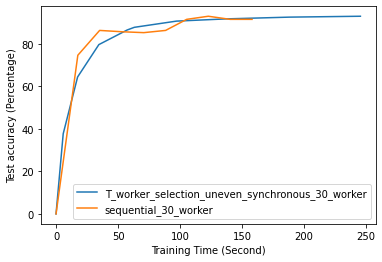}
  \caption{MINST 30 workers]}
\end{subfigure}
\begin{subfigure}{.5\textwidth}
  \centering
  \includegraphics[width=0.8\linewidth, height=4cm]{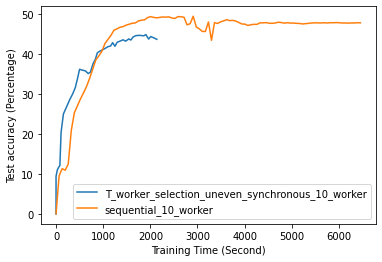}
  \caption{CIFAR 10 workers}
\end{subfigure}
\begin{subfigure}{.5\textwidth}
  \centering
  \includegraphics[width=0.8\linewidth, height=4cm]{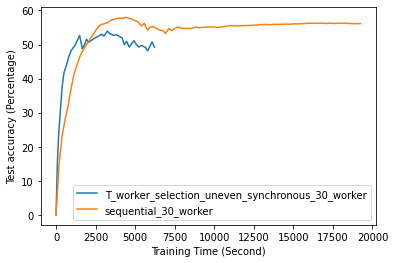}
  \caption{CIFAR 30 workers}
\end{subfigure}
\caption{Algorithm \ref{algo:tbased} Worker Selection (synchronous) VS Sequential}
\label{fig:syn_t_ws_vs_s}
\end{figure*}
\begin{figure*}[!ht]
\begin{subfigure}{.5\textwidth}
  \centering
  \includegraphics[width=0.8\linewidth, height=4cm]{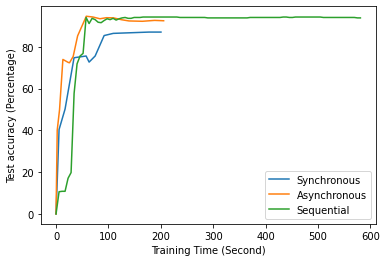}
  \caption{MINST 10 workers]}
\end{subfigure}%
\begin{subfigure}{.5\textwidth}
  \centering
  \includegraphics[width=0.8\linewidth, height=4cm]{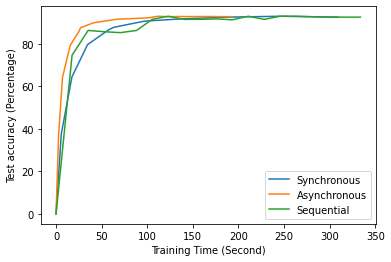}
  \caption{MINST 30 workers]}
\end{subfigure}
\begin{subfigure}{.5\textwidth}
  \centering
  \includegraphics[width=0.8\linewidth, height=4cm]{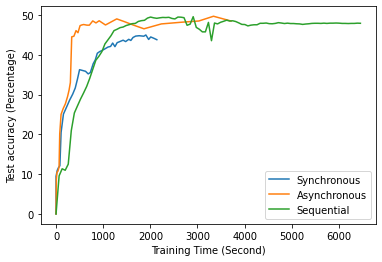}
  \caption{CIFAR 10 workers}
\end{subfigure}
\begin{subfigure}{.5\textwidth}
  \centering
  \includegraphics[width=0.8\linewidth, height=4cm]{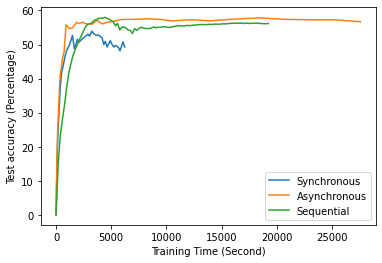}
  \caption{CIFAR 30 workers}
\end{subfigure}
\caption{Algorithm \ref{algo:tbased} Worker Selection (Synchronous) VS (Asynchronous) VS Sequential}
\label{fig:asyn_vs_syn_vs_s}
\end{figure*}

Fig.~\ref{fig:sequential_vs_even} to Fig.~\ref{fig:asyn_vs_syn_vs_s} illustrate the results of FL training under configurations in tables~\ref{table:ten_worker} and \ref{table:thirty_worker}, which are described in the following.
\par
FL with even data distribution requires less time to reach stable accuracy, meaning faster training at the initial stage. But sequential one eventually reaches a better accuracy. This can be seen from Fig.~\ref{fig:sequential_vs_even} that the even data distribution line reaches a high accuracy before the sequential training. However, sequential training reaches better results over time. Moreover, as it can be seen from Fig.~\ref{fig:even_vs_uneven}, the time required for sets of workers that have even or uneven amounts of training data to reach stable accuracy is similar.
\par
According to Fig.~\ref{fig:rws_vs_sequential}, random worker selection eventually reaches the same accuracy level as sequential implementation. However, random worker selection requires a longer time to get the same accuracy as sequential. Moreover, it shows that the accuracy growth of random worker selection is unstable compared to sequential one. Besides, based on Fig.~\ref{fig:rminmax_vs_sequential}, the r-min and r-max worker selection algorithm is not more time efficient compared to sequential implementation. Moreover, incorrect initialization of $rmin$ and $rmax$ can lead to inefficient training processes such that accuracy never approaches the potential accuracy achievable based on all data from all workers. This latter case can be seen from Fig.~ \ref{fig:r-min-r-max-initialisation} that when $rmax$ is initialized to 5, 6, and 7, the accuracy stays around 15\%. In contrast, theoretical accuracy is around 50\%. Overall, the Algorithm~\ref{algo:rminmax} is time inefficient because $rmin$ and $rmax$ diverge too fast in the early stage of ML training. Since model weights are randomly initialized, there is significant accuracy growth in the earlier rounds of training. This accelerates the update of $rmin$ and $rmax$ based on the update Eq.~\ref{equation:rmin} and causes faster divergence. When $rmin$ and $rmax$ becomes far different, slow computing workers are also considered since the time required to complete minimal training is much less than the time necessary for faster workers to meet maximally allowed training epochs. Thus, the worker selection algorithm \ref{algo:rminmax} quickly turns into selecting all workers after a few rounds of aggregation on the server.
\par

According to Fig.~\ref{fig:syn_t_ws_vs_s}, the worker selection algorithm \ref{algo:tbased}, when combined with synchronous FL training, outperforms random worker selection and sequential ML training during the early phase of training. This shows the effectiveness of the worker selection algorithm \ref{algo:tbased}. However, such performance only takes part in an early phase of training, while sequential training always reaches stable accuracy faster. The worker selection algorithm \ref{algo:tbased} is more time efficient in the early phase compared to sequential ML training because only fast computing workers are selected to participate in training. When the FL training requires slower workers to join in later rounds of training, since synchronous FL requires faster workers to wait for slower workers, it becomes time inefficient as training progress to later rounds. This is also proved by the fact that the accuracy of sequential training exceeds it in later training rounds. As shown in Fig.~\ref{fig:asyn_vs_syn_vs_s}, the worker selection algorithm \ref{algo:tbased} combined with asynchronous FL training has similar performance in the earlier phase. However, during a later stage of training, asynchronous FL has faster accuracy growth. This makes asynchronous FL and worker selection algorithm \ref{algo:tbased} more time efficient than synchronous FL or sequential ML training. Asynchronous FL, even when it involves slower workers, does not require faster workers to wait, resulting in asynchronous FL outperforms sequential and synchronous training in terms of time efficiency.
\section{Conclusions and Future Work}
\label{conclusion}
In this paper, we have designed and implemented a lightweight and containerized framework for FL, called FLight, by extending the FogBus2 framework. FLight enables the integration of new ML models and supports different mechanisms relating to worker selection access control and storage implementation to be extended easily. Moreover, two worker selection algorithms are introduced in this work to improve the training time efficiency. FLight enables easy extension of mechanism related to FL. New ML models can be extended as long as import and export model weights and merging model weights are defined. Moreover, the worker selection strategy can be extended by overriding the worker selection function skeleton, where multiple system parameters relating to available workers are available for different algorithm designs. Also, the flexible storage model design enables the model weights to be stored on different media, which supports deployment on various computing resources. The implementation also supports asynchronous FL. The easy extension and lightweight property of the FLight makes it a good tool for comparing different federated learning mechanism design. 
In future works, we plan to integrate other the-state-of-the-art FL optimization mechanisms into the FLight framework. Moreover, the current framework provides the opportunity to integrate other ML techniques in a distributed manner, such as some of the state-of-the-art distributed deep reinforcement learning techniques for dynamic scheduling of resources \cite{goudarzi2021distributedDDRL}. Also, considering security and privacy perspectives, although FL inherently protects data privacy as data stays locally at its origination, model weights shared to remote sites still can leak information about training data. Thus, extra modification on shared model weights is necessary to prevent any sensitive information regarding training data leaked out through the format of ML model weights. This is important since protecting training data privacy is the key goal of FL which emphasize this feature should be integrated into the Flight framework. Finally, because Flight is a containerized FL framework, one logical step to expand its properties is enabling container orchestration properties for the Flight framework. So, we plan to deploy the Flight on container orchestration platforms, such as Kubernetes and K3S, to offer higher scalability, automated monitoring, and failure handling for the Flight framework. 




%





\ifCLASSOPTIONcaptionsoff
  \newpage
\fi




\section*{Software Availability}
The source code of the FLight framework is accessible from:
\href{https://github.com/Cloudslab/FLight}{https://github.com/Cloudslab/FLight}

\bibliographystyle{IEEEtran}
\bibliography{IEEEabrv,Bibliography}

\vfill


\end{document}